\documentclass[a4paper,twocolumn,superscriptaddress,longbibliography]{revtex4-1}

\usepackage[utf8]{inputenc}
\usepackage{braket}
\usepackage{amsmath}
\usepackage{mathtools}
\usepackage{amssymb}
\usepackage{amsfonts}
\usepackage{graphicx}
\usepackage{hyperref}
\usepackage{comment}
\DeclareMathOperator{\Tr}{Tr}
\newcommand{\tp}{\mathsf{T}}
\newcommand{\avg}[1]{\langle #1 \rangle}
\newcommand{\Id}{\mathbb{I}}
\newcommand{\tavg}{{\ensuremath{t\text{-avg}}}}
\newcommand{\dyad}[2]{\ket{#1}\!\bra{#2}}
\newcommand{\gammagamma}{\left(\begin{smallmatrix} \gamma & \\ & \gamma \end{smallmatrix}\right)}
\newcommand{\gammazero}{\left(\begin{smallmatrix} \gamma & \\ & 0 \end{smallmatrix}\right)}
\newcommand{\zerogamma}{\left(\begin{smallmatrix} 0 & \\ & \gamma \end{smallmatrix}\right)}

\begin{document}

\title{Localization with random time-periodic quantum circuits}

\date{\today}

\author{Christoph Sünderhauf}
\affiliation{Max-Planck-Institut für Quantenoptik, Hans-Kopfermann-Str.~1, 85748 Garching, Germany}
\author{David Pérez-García}
\affiliation{Departamento de Análisis Matemático, Universidad Complutense de Madrid, Plaza de Ciencias~3, 28040 Madrid, Spain}
\affiliation{ICMAT, Nicolas Cabrera, Campus de Cantoblanco, 28049 Madrid, Spain}
\author{David A.~Huse}
\affiliation{Physics Department, Princeton University, Princeton, New Jersey 08544, USA}
\author{Norbert Schuch}
\affiliation{Max-Planck-Institut für Quantenoptik, Hans-Kopfermann-Str.~1, 85748 Garching, Germany}
\author{J.~Ignacio Cirac}
\affiliation{Max-Planck-Institut für Quantenoptik, Hans-Kopfermann-Str.~1, 85748 Garching, Germany}

\begin{abstract}
 We consider a random time evolution operator composed of a circuit of random unitaries coupling even and odd neighboring spins on a chain in turn. In spirit of Floquet evolution, the circuit is time-periodic; each timestep is repeated with the same random instances. We obtain analytical results for arbitrary local Hilbert space dimension $d$: On a single site, average time evolution acts as a depolarising channel. In the spin 1/2 ($d=2$) case, this is further quantified numerically. For that, we develop a new numerical method that reduces complexity by an exponential factor. Haar-distributed unitaries lead to full depolarization after many timesteps, i.e. local thermalization. A unitary probability distribution with tunable coupling strength allows us to observe a many-body localization transition.
 In addition to a spin chain under a unitary circuit, we consider the analogous problem with Gaussian circuits. We can make stronger statements about the entire covariance matrix instead of single sites only, and find that the dynamics is localising. For a random time evolution operator homogeneous in space, however, the system delocalizes.
 
\end{abstract}

\maketitle

\tableofcontents

\section{Introduction}

The dynamics of many-body quantum systems has revived the interest in thermalization and localization. In closed systems, there are states that do not thermalize. A simple example is a single particle in a random potential that is Anderson localized \cite{Anderson}. But even if one includes interactions, a new way of many-body localization (MBL) can emerge that also prevents thermalization \cite{Bloch_MBL}. Despite great progress in understanding MBL during the last years (see eg.~\cite{Huse_MBL_l-bits} or the review \cite{Abanin_MBL_review}), there are still many open questions.

A typical scenario studied in the context of localization is a system on a one-dimensional lattice, with a short-ranged Hamiltonian containing a kinetic term and a random potential for each site. In the absence of interactions, this single-particle problem displays Anderson localization:
Starting in one position, the probability of finding the particle at the same position after arbitrary time is lower bounded, and the probability for other positions is exponentially suppressed \cite{Stolz, Aizenman}. If one adds interactions, the system can find itself in the thermal or MBL phase, usually dependent on disorder strength. Starting with some information in a specific position, in the thermal phase it will flow away and cannot be recovered locally, and in the MBL phase there will still be traces present at the same position after arbitrarily long times, despite some information slowly flowing away \cite{Pollmann_entanglement_growth_MBL,banuls2017dynamics,znidaricPRB2008}.

Another scenario is so-called Floquet evolution. There, one considers not continuous time evolution generated by time-invariant Hamiltonians, but a discrete-time evolution operator repeated for subsequent timesteps. It may arise from a periodic drive or be directly given as a unitary model. Floquet systems are a formidable setting to study localization, because even energy ceases to be a conserved quantity. Anderson localization has been proven for specific Floquet systems \cite{Hamza}. It has been found that Floquet systems are compelling examples for MBL \cite{PRL-Abanin-MBL_in_per_driven_sys,PRL-Moessner-Fate_MBL_per_driving} which yield sharper transitions between thermal and MBL phases \cite{Huse_Floquet}.

In addition, circuits of random unitaries have recently been used as a model of chaotic systems \cite{Tibor1, Tibor2, Amos_Andrea_1, like_Tibor1, like_Tibor2, Nahum_entanglement_growth,Huse_random_circuit, Nahum2018emergent_statmech, emerson2003pseudo}.
In \cite{Tibor1, like_Tibor1}, time evolution by a unitary circuit of fixed geometry but independently Haar-distributed random gates at each time step was studied. That model exhibits thermalization to an infinite temperature state, and the authors found ballistic spreading of quantum information by considering the out-of-time-ordered correlator. Subsequently the model was extended to a similar setup \cite{Tibor2, like_Tibor2} with a conservation law. In \cite{Amos_Andrea_1}, the authors consider the same unitary circuit in a Floquet setting, where subsequent timesteps are repeated with the same random instances. In the limit of infinite local Hilbert space dimension for each qudit, they find thermalization to an infinite temperature state and calculate several values like the spectral form factor or the exponentials of some Renyi-entropies.
In other related work \cite{Shenker_OnsetRMT,kos2017rmt,Prosen2018exact}, thermalization of spin chains for certain continuous-time dynamics was found in the context of the average spectral form factor.

Here, we consider several variations of Floquet evolution with a unitary circuit, and analyse if there is localization. We consider as time evolution operator a quantum circuit of depth two, which consists of two alternating layers of random nearest-neighbor unitaries coupling even and odd pairs of sites in turn. The two layers are repeated identically for subsequent timesteps such that the total circuit is periodic in time, in the spirit of Floquet evolution. This circuit geometry is the discrete analogue of local time-independent Hamiltonian evolution (and could also be obtained by a Trotter decomposition, or the standard form of an index zero matrix product unitary {\csname @fileswfalse\endcsname\cite{MPU}}, for example).
We perform an average within a (sub)set of unitaries.
Typically, we start with a completely mixed state everywhere and a pure state at one site and look at the reduced state of that and other sites at some later time, and determine whether it depends on the initial state, corresponding to localization.

The scenarios we consider are the following: (A) Gaussian circuits, acting on fermionic chains with one mode per site and Gaussian evolution, where the nearest-neighbor unitaries in the circuit are operations that stay within the manifold of fermionic Gaussian states. (B) Spins, with a qudit per site and arbitrary constituent unitaries in the circuit.
The first scenario, (A) Gaussian circuits, extends the typical situation in Anderson localization, since particle number is not conserved. In this scenario, we consider inhomogeneous as well as homogeneous Floquet circuits, where the unitaries coupling sites are independently random for each pair of neighbors or the same along the entire chain. We find that the inhomogeneous setting exhibits localization, whereas the homogeneous Floquet circuit leads to delocalization.

The second scenario, (B) spins, is similar to the models studied in \cite{Amos_Andrea_1,Tibor1,like_Tibor1}. In contrast to \cite{Amos_Andrea_1}, in our work the local Hilbert space dimension of each spin is finite, and in contrast to \cite{Tibor1, like_Tibor1}, we work in a Floquet setting. We prove that on a single site, the time evolution acts as a depolarising channel. Further, we find that a chain of qubits can exhibit thermalization or MBL, depending on the probability distribution used to average the unitaries in the circuit; we observe the corresponding phase transition.

Our setup is difficult computationally and analytically, because it requires to study dynamics of many-body systems, averaged over instances of the random Floquet circuit. Methods to exactly calculate averages \cite{Weingarten_graphical, Brouwer_Beenakker} work well when each random matrix appears a small amount of times, or for large dimensions where asymptotic behaviour is available. These methods are not useful in our setting, since the same random matrices reappear in each timestep (contrary to \cite{Tibor1,like_Tibor1}) and we have a fixed finite dimension of the spins (contrary to \cite{Amos_Andrea_1}). Instead, we derive analytical results in both cases with a technique we call the twirling technique. It is based on a property of the average, which basically allows us to move arbitrary single-site unitaries through the quantum circuit such that they only appear twice, at the beginning and end, relating initial and final states.

Apart from that, we also perform numerical calculations {\csname @fileswfalse\endcsname\cite{Edelman_Rao_RMT}} in both cases. For (A) Gaussian circuits, we can work with the covariance matrix formalism, which is very efficient and allows us to explore very large systems. For (B) spins, the Hilbert space is exponential in chain length. We develop a new numerical method which combines tensor networks and Monte Carlo ideas, drawing from simplifications provided by the analytic results. It reduces the memory and time complexity from $2^{4t}$ to $2^t$ for $t$ timesteps. This allows us to study relativity long times which, in turn, enables the simulation of up to 39 spins.

This article is organized as follows. First, we introduce the precise models in section~\ref{sec:settings} and the quantities we will compute. In section~\ref{sec:results} we present the main results of this work, and leave the derivations for section~\ref{sec:proofs}. There, we also present the twirling technique (section~\ref{sec:twirling technique}) used throughout the paper, which can also be of interest on its own.   Finally, in section~\ref{sec:unitary details numeric} we present the new numerical method used for spin chains.

\section{Settings \& Questions}
\label{sec:settings}

\begin{figure}
 \centering
 \includegraphics[width=4cm]{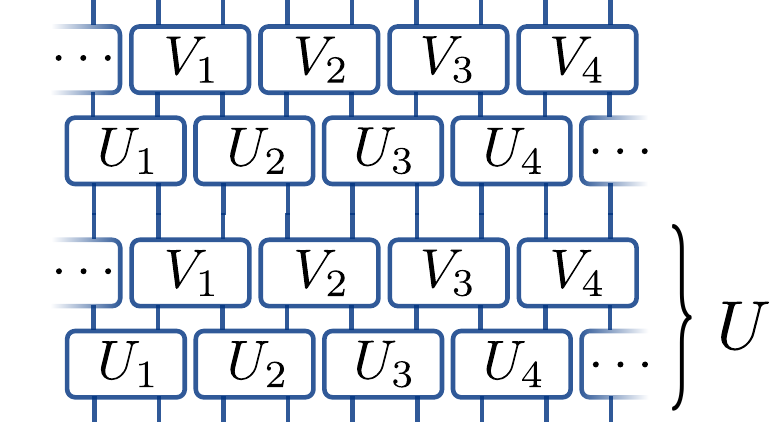}
 \caption{Random time evolution operator $U^2$ for two timesteps. The vertical lines indicate sites of the chain; each unitary couples two neighboring sites. In the spirit of Floquet evolution, the total evolution operator is time-periodic; the time step is repeated with the same random instances of $U_i,V_i$.}
 \label{fig:timestep}
\end{figure}

For a one-dimensional chain of $N$ particles, we consider a random unitary time evolution operator $U$ composed of random nearest-neighbor gates according to some probability distribution. The time evolution operator is the unitary circuit with the fixed geometry sketched in Fig.~\ref{fig:timestep} and can be written as
\begin{equation}
 U = \left(\bigotimes_i V_i\right)\left(\bigotimes_i U_i\right).
 \label{eq:unitary time evolution operator}
\end{equation}
The unitary $U_i$ acts on particles $2i-1$ and $2i$ while $V_i$ acts on sites $2i$ and $2i+1$. These two layers are repeated identically (with the same random instances of $U_i,V_i$) in spirit of Floquet evolution, in contrast to other models \cite{Tibor1, like_Tibor1} where each timestep is different.

In this article, we study the random circuit as a time evolution operator that is a (A) Gaussian circuit for fermionic chains or (B) Unitary circuit for spin chains. Throughout, the average $\avg{\cdot}$ denotes averaging over $U_i,V_i$. In the next two subsections, we give details of both settings, and define the probability distributions used for the average $\avg{\cdot}$ in either setting.

\subsection{Gaussian circuits}

First, we consider the problem for a chain of fermionic systems with one fermionic mode per site. Each of the $N$ modes has two Majorana operators 
\begin{equation}
c_{2n-1}=a_n^\dagger + a_n,\ c_{2n} = -i(a_n^\dagger-a_n),
\end{equation}
with the creation/annihilation operators $a_n^\dagger$/$a_n$.
The two-point correlation functions of Majorana operators for each fermionic state $\rho$ can be gathered in the covariance matrix
\begin{equation}
 \Gamma_{kl} := \frac{i}{2}\Tr(\rho[c_k,c_l]).
\end{equation}
Each site of the chain corresponds to a $2\times2$ block because each site is associated with two Majorana operators. A fermionic Gaussian state (i.e.~those that can be generated by the vacuum of $a_n$ by Gaussian functions of the Majorana operators) is completely and uniquely characterized by its covariance matrix. Here, we consider the covariance matrices of not only Gaussian but arbitrary initial states with vanishing two-point correlators at non-zero distances.

We build the Gaussian circuit of transformations that map Gaussian states to Gaussian states (but can still be applied to general states). The most general such unitary operation acts on the covariance matrix by an orthogonal transformation $O\in O(2N)$, specifically $\Gamma \to O\Gamma O^\tp$.

We will consider two classes of these unitary transformations: Gaussian operations generated by Hamiltonians quadratic in the Majorana operators, which correspond to special orthogonal transformations $O\in SO(2N)$ in the covariance matrix formalism \cite{Matchgates,Bravyi}, and the larger class of all operations $O\in O(2N)$ which includes local particle-hole transformations
\footnote{%
This class includes (local) particle hole transformations. For example for a single fermionic mode, particle-hole transformation corresponds to the unitary $U = a + a^\dagger$ ($a^\dagger/a$ creation/annihilation operators) and in the covariance matrix formalism, to $\left(\begin{smallmatrix}1&0 \\ 0 &-1\end{smallmatrix}\right)\in O(2)$ with negative determinant. All of the transformations we consider have definite parity as required by superselection rules.}.
Subsequently, we only consider the covariance matrices of initial and final states.

In this setup, the unitary-circuit time evolution operator (\ref{eq:unitary time evolution operator}) is represented as an orthogonal transformation $O\in O(2N)$ built of random two-site operations $P_i,Q_i\in O(4)$. With periodic boundary conditions,
\begin{equation}
 O = G \left( \bigoplus_{i=1}^{N/2} Q_i \right) G^T \left(\bigoplus_{i=1}^{N/2} P_i\right),
 \label{eq:fermion evolution operator}
\end{equation}
where 
\begin{equation}
 G = \begin{pmatrix}0 & & & \Id_2 \\ \Id_2 & 0 & &  \\ & \ddots & \ddots & \\ & & \Id_2 & 0  \end{pmatrix}
\end{equation}
takes care of circularly shifting $\bigoplus Q_i$ by one site; i.e.~two matrix elements down and right. Thereby $P_i$ couples site $2i-1$ with $2i$ and $Q_i$ couples site $2i$ with $2i+1$.

Our quantity of interest is the average final state $\avg{\Gamma_t}$ after $t$ timesteps of an initially uncorrelated product state $\Gamma_0$, i.e.~with a $2\times2$ block-diagonal covariance matrix. In this formalism its covariance matrix is
\begin{equation}
 \avg{\Gamma_t} = \avg{ O^t \Gamma_0 O^{t\dagger} }.
 \label{eq:fermion final state def}
\end{equation}
For the expectation value $\avg{\cdot}$, we consider two probability measures for the $P_i,Q_i$: the Haar measure for the orthogonal group $P_i,Q_i\in O(4)$ and the Haar measure for the special orthogonal group $P_i,Q_i\in SO(4)$.
The Haar distribution (see eg.~\cite{Haar_measure}) for the orthogonal (special orthogonal) group $O(4)$ ($SO(4)$) is defined as the unique distribution with the property of Haar invariance, which mandates that any transformation
\begin{equation}
 P \to A P B, \text{\ for any\ } A,B\in O(4)\ (SO(4))
\end{equation}
does not affect averages $\avg{\cdot}$ with respect to $P\in O(4)\ (SO(4))$
\footnote{The Haar distribution will allow us to derive some analytical results. It treats all bases on an equal footing (this means there is no preferred local basis for the evolution), and rotation angles are random. It is also the most depolarising measure and thus one would expect to obtain the most extreme results. Physically, it corresponds to having magnetic fields not only with a random strength, but also a random direction.}.
The long-time behaviour of an initial covariance matrix is readily accessible to numerical calculations even on long chains, because we need only operate on its covariance matrix, whose dimension grows merely linearly in system size.

We consider two scenarios. In the first scenario, all $P_i,Q_i$ are independently distributed according to one of the Haar measures. This situation is related but not equivalent to that studied in context of Anderson localization. The main reason is that the average over $O$ includes transformations $P_i,Q_i$ that do not conserve particle number. Thus, a question to be addressed is whether the well-studied phenomenon of Anderson localization still exists, or if it is modified. To this end we ask, does the average final state $\avg{\Gamma_t}$
contain remnant information about the initial state $\Gamma_0$?
The corresponding results are reported in section~\ref{sec:results fermions inhomogeneous}.

Furthermore, we study a second scenario, the homogeneous setting where the time evolution operator $O$ is 2-site-translation invariant. In that scenario, randomness is the same for all sites, $P_i=P_j$ and $Q_i=Q_j$, such that there are only two independent transformations; here we consider only the Haar measure over $O(4)$. Again, we ask the same question: Does an impurity in an otherwise translation-invariant state spread all over the chain or stay localized? We present the answer in section~\ref{sec:fermions results homogeneous}.
Occasionally, the time average
\begin{equation}
 \avg{\Gamma_\tavg} := \lim_{T\to\infty} \frac{1}{T} \sum_{t=0}^{T-1} \avg{\Gamma_t}
\end{equation}
is used to assess the localising or delocalising properties. Physically, it captures the long-time behaviour of a typical state. The additional average allows us to make stronger statements.

\subsection{Spins}
\label{sec:settings spins}
After studying the evolution of a chain of fermions under Gaussian circuits, we turn to a chain of interacting spins. 
All particles along the chain have a local Hilbert space dimension $d$, which may be arbitrary.
In that setting, all of the unitaries $U_i,V_i$ composing the circuit $U$ are general unitaries of $U(d^2)$, independently distributed according to some probability distribution for the average $\avg{\cdot}$.

We will consider different probability distributions for $U_i,V_i\in U(d^2)$ with the common property of single-site Haar invariance. This means that any transformation of a $U_i$ or $V_i$ of the form 
\begin{equation}
U_i \leftrightarrow (w_1\otimes w_2)U_i(w_3\otimes w_4)
\end{equation}
does not affect averages $\avg{\cdot}$, for arbitrary choice of $w_j\in U(d)$. For example, the unitary Haar distribution on $U(d^2)$ has this property. It is a distribution uniquely defined by Haar invariance (see eg.~\cite{Haar_measure}), which means that transformations of the form $U\to A U B$, for arbitrary $A,B\in U(d^2)$, do not affect any averages with respect to the Haar distribution of $U\in U(d^2)$ \cite{Note2}.

In this article, we characterize the average final state $\avg{\rho_t}$ obtained from an initial density matrix $\rho_0$ after $t$ timesteps:
\begin{equation}
 \avg{\rho_t} := \avg{U^t\rho_0 U^{t\dagger}}.
 \label{eq:average time evolution}
\end{equation}
In particular, for arbitrary initial states $\rho_0$, we will find a relation between the reduced initial state $\rho_0^\text{red} := \Tr_{\{1\ldots N\}\backslash\{n\}}\rho_0$ on a single site $n$ and the reduced state $\avg{\rho_t^\text{red}} := \Tr_{\{1\ldots N\}\backslash\{n\}}\avg{\rho_t}$ of the average final state on the same site.
We find that on a single site, average time evolution acts as a depolarising channel. This result is formulated in section~\ref{sec: results spins analytical}.

With this local characterization of initial and average final states, we assess the long-time behaviour of $\avg{\rho^\text{red}_t}$ numerically. Interacting systems may thermalize, or else display many-body localization. In this context, we ask, does $\avg{\rho_t}$ locally remember the initial state (localization) or not (thermalization)? For example, imagine an initial state that is homogeneous except for an impurity at one site. Then we ask, after average time evolution, can we perform local measurements at the same or other sites to recover information about the position and initial state of this impurity? We present our corresponding results in sections~\ref{sec: results spins haar} and~\ref{sec: results spins MBL}.

\section{Results}
\label{sec:results}

In this section, we present our main results for (A) Gaussian circuits or (B) spins. We leave the details of the derivations, as well as the methods used to obtain them, for the next sections.

\subsection{Gaussian circuits}

First, we consider the setting of Gaussian circuits. We will first consider the inhomogeneous case, where orthogonal matrices for different sites are independently random. Then we will give the results for the homogeneous case, where the time evolution operator is invariant under translations by two sites.

\subsubsection{Inhomogeneous evolution exhibits localization}
\label{sec:results fermions inhomogeneous}

For uncorrelated initial states $\Gamma_0$, i.e.~$2\times2$ block-diagonal $\Gamma_0$, we find the following result:
\begin{equation}
 \label{eq:fermion inhomogeneous result}
 \avg{\Gamma_t} = c(t,N) \Gamma_0.
\end{equation}
The constant $c(t,N)$ is independent of the initial state.  We obtain this result for both the orthogonal Haar measure, $P_i,Q_i\in O(4)$, as well as the special orthogonal Haar measure, $P_i,Q_i\in SO(4)$, with the same constant $c(t,N)$ in both cases. The latter case holds as long as $t < (N-1)/2$, i.e.~the system is large enough to accommodate the lightcone without self-intersections.
Hence in the thermodynamic limit, $O(4)$ and $SO(4)$ Haar averages are equivalent in this setting. We prove these results in section~\ref{sec:fermions details inhomogeneous}.

We further study $c(t,N)$ numerically, and plot it in Fig.~\ref{fig:fermion c(t,N)} as a function of time steps $t$ for different system sizes $N$. We observe that $c(t,N)$ converges to a fixed value $c\approx0.06$, irrespective of $N$. 
After one time step, $c(1,N)=0$ exactly, which simply is thermalizing evolution with independent Haar distributed orthogonals. Only at longer times does the time-periodic structure of the circuit become manifest and result in appreciably change in measure.
Since $c(t,N)$ reaches a non-zero value, we find that Anderson localization still happens in this extended setup; an initially localized impurity stays localized. Each site of the initial state is simply scaled towards the thermal mixture $\Gamma=0$ by the same factor $c(t,N)$. Nevertheless, after average time evolution, the initial state's covariance matrix can still be fully reconstructed from measured expectation values, albeit their variances increase.

\begin{figure}
  \centering
  \includegraphics[width=8cm]{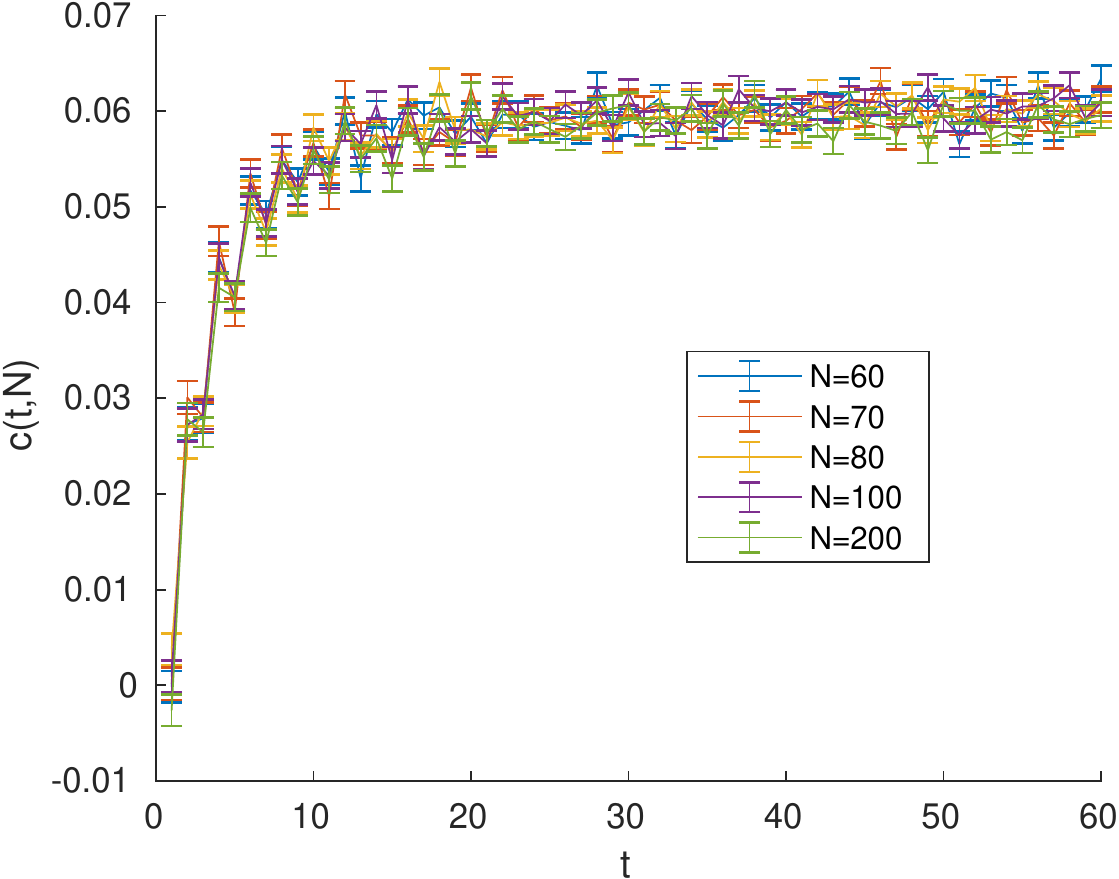}
  \caption{The constant $c(t,N)$ with which a fermionic covariance matrix is scaled by random time evolution for $t$ timesteps, see eq.~(\ref{eq:fermion inhomogeneous result}). We generate $10^4$ samples of $O(4)$-Haar-distributed $P_i,Q_i\in O(4)$ for each data point. Surprisingly, $c(t,N)$ does not depend on system size $N$ even for $t$ large enough such that the lightcone wraps around the periodic boundaries. }
  \label{fig:fermion c(t,N)}
\end{figure}

\subsubsection{Homogeneous evolution delocalizes}
\label{sec:fermions results homogeneous}

Next, let us consider a homogeneous time evolution operator, where $P_i=P_j \in O(4)$ and $Q_i = Q_j\in O(4)$ are distributed according to the orthogonal Haar measure.
Let $\Gamma_0^n$ be an initial state with a single site $n$ occupied and all others maximally mixed. This is a zero matrix, except that the $2\times2$ block for site $n$ is $\gamma := \left(\begin{smallmatrix}0&1\\-1 & 0\end{smallmatrix}\right)$.
In section~\ref{sec:fermions details homogeneous} we show the time-averaged final state
of this initially localized state to have the covariance matrix
\begin{equation}
 \avg{\Gamma_\tavg^n} := \lim_{T\to\infty}\frac{1}{T}\sum_{t=0}^{T-1} \avg{\Gamma_t^n} = \frac{1}{N/2}\Gamma_\star,
 \label{eq:fermions homogeneous result}
\end{equation}
under a plausible assumption about disjointness of spectra of matrices that are multiplied by Haar-random orthogonal matrices which we also verified numerically. We characterize $\Gamma_\star$ further in section~\ref{sec:fermions details homogeneous}.

An important part of the result is that the covariance matrix $\Gamma_\star$ depends not on the precise value of $n$ but only on $n\!\!\mod4$. Thus, the location $n$ of the impurity cannot be reconstructed from $\avg{\Gamma_\tavg^n}$. Moreover, in the thermodynamic limit, the prefactor $1/(N/2)$ causes $\avg{\Gamma_\tavg^n}$ to reach the infinite temperature thermal mixture 0.
In conclusion, our result implies the absence of localization.

A complementary viewpoint of delocalization is provided by the delocalization of eigenvectors of a single generic random instance of the time evolution operator. In section~\ref{sec:fermions eigenvector delocalization} we prove how this allows us to bound all matrix elements of $\Gamma_\tavg^n$ for a generic evolution operator $O$ with non-degenerate spectrum:
\begin{equation}
 |(\Gamma_\tavg^n)_{ij}| \le \frac{16}{N} \to 0
 \label{eq:fermion results eigenvector delocalization}
\end{equation}
in the thermodynamic limit, without resorting to an ensemble average $\avg{\cdot}$. On the one hand, this result is stronger than (\ref{eq:fermions homogeneous result}) insofar as it shows $\Gamma_\tavg^n\to0$ in the thermodynamic limit already for single instances of the time evolution operator. On the other hand, it only gives a bound $\le \frac{16}{N}$ and not an explicit form.

\subsection{Spins}

We now move from Gaussian circuits to interacting spins. The average $\avg{\cdot}$ is now an average over all nearest-neighbor unitaries $U_i,V_i\in U(d^2)$ comprising the time evolution operator, independently distributed according to some probability distribution with single-site Haar invariance (see~\ref{sec:settings spins}).
Here we will first present the statement that relates the evolution of a single site with a depolarising channel. Then, we show results which indicate the absence of localization when averaging with the Haar measure on $U(4)$. Finally, we will consider different unitary ensembles, which vary in the degree of entanglement the $U_i,V_i$ generate and present numerical evidence for a thermal-MBL phase transition.

\subsubsection{Depolarising channel on each site}
\label{sec: results spins analytical}

Our first result is, that on a single site, the average time evolution (\ref{eq:average time evolution}) acts as a depolarising channel. To make this result precise, consider an arbitrary initial state $\rho_0$. Split its reduced density matrix for one site
\begin{equation}
 \rho_0^\text{red} = \Id_d/d + \bar{\rho}_0^\text{red}
\end{equation}
into traceful and traceless part $\bar{\rho}_0^\text{red}$. For the evolved reduced state at the same site we prove
\begin{equation}
 \avg{\rho_t^\text{red}} = \Id_d/d + \alpha(t)\,\bar{\rho}_0^\text{red}.
 \label{eq:unitary depolarising channel}
\end{equation}
This corresponds to a depolarising channel \cite{Nielsen_Chuang} with depolarization probability $1-\alpha(t)$.
The real constant $\alpha(t)$ is independent of the initial state. Provided the lightcone ($2t+1$ sites in width) around the site fits into the system, it is also independent of the position of the site and of system size. Moreover, it is striking that the final state on a single site is affected only by the initial state on the same site, and is independent of the initial state at all other sites. We prove (\ref{eq:unitary depolarising channel}) in section~\ref{sec:unitary details analytic} where we also derive a similar formula for the two-site reduced density matrix.

If the initial state is free of inter-site correlations, with all but one site completely mixed,  the final state can be fully characterized. Thereby the initial state $\rho_0 = \Id_d/d\otimes\rho_0^\text{red}\otimes\Id_d/d\otimes\Id_d/d\otimes\cdots$ evolves to a final state with the same structure $\avg{\rho_t} = \Id_d/d\otimes\avg{\rho_t^\text{red}}\otimes\Id_d/d\otimes\Id_d/d\otimes\cdots$.

To understand the behaviour of the system, it is necessary to determine the behaviour of $\alpha(t)$.
For this, we will study $\alpha(t)$ numerically for spin 1/2 particles, $d=2$. 
In order to access long times, we use a new numerical method (section~\ref{sec:unitary details numeric}). It reduces the complexity for $t$ timesteps from $2^{2(2t+1)}$ to $2^t$ and uses an importance sampling technique to lower the variance. Since the number of spins involved after $t$ timesteps is $2t+1$, this in turn has allowed us to reach $39$ of them while maintaining an effectively infinite system size.

\subsubsection{Haar-distributed unitaries thermalize}
\label{sec: results spins haar}
\begin{figure}
 \centering
 \includegraphics[width=\columnwidth]{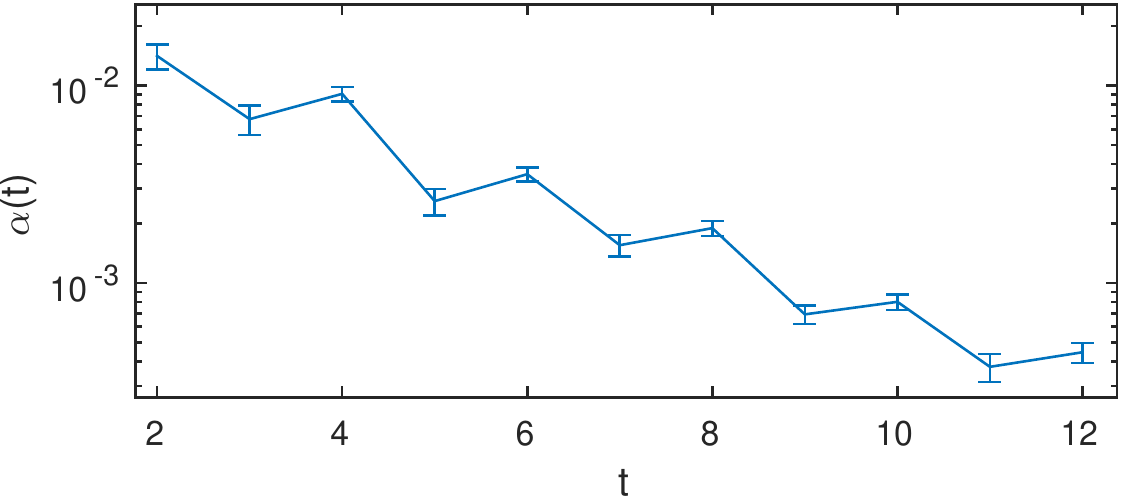}
 \caption{The constant $\alpha(t)$ relating initial and final states (\ref{eq:unitary depolarising channel}) for Haar-distributed unitaries. The figure is indicative of thermalization at long times as $\alpha$ vanishes exponentially in time. For each data point, $10^3$ samples of random unitaries were generated.}
 \label{fig:unitary haar alpha}
\end{figure}

As a concrete probability distribution for the unitaries $U_i,V_i$, we first consider the Haar distribution on $U(4)$.
In Fig~\ref{fig:unitary haar alpha} we present numerical results for this probability distribution. They show that $\alpha$ vanishes exponentially as $t\to\infty$, with a half-life of about 1.8 timesteps. Therefore we find thermalization to a locally infinite temperature state: The map (\ref{eq:unitary depolarising channel}) describing a single site's evolution becomes completely depolarising in the limit $t\to\infty$ where $\alpha\to0$. A similar result has been obtained in \cite{Shenker_OnsetRMT} in a Hamiltonian (continuous time evolution) setting.

This result is in stark contrast to the analogous setting with Gaussian circuits (section~\ref{sec:results fermions inhomogeneous}). A Floquet operator built of unitaries conserving Gaussianity as studied in that setting causes localization, while taking into account all unitaries, it causes thermalization. The reason for this difference can be attributed to the fact that MBL phases are not ubiquitous in parameter space \cite{Nandkishore_Huse_Review}, whereas Anderson localization is (in 1D models, as analysed here).

\subsubsection{Tunable coupling strength and MBL transition}
\label{sec: results spins MBL}
As seen in the previous section, the Haar distribution exhibits thermalising behaviour, since MBL can  typically only be found for strong random potentials relative to the coupling \cite{{Nandkishore_Huse_Review}}. In practice, Haar-distributed $U_i$ and $V_i$ contain many highly entangling operators which can move information that is initially contained in one site across the chain. This opens up the question of whether MBL can be found by considering less entangling operations. We therefore modify the distribution used for the unitaries composing the time evolution operator.

Every unitary in $U(4)$ can be cast in the form \cite{unitary_decomposition}
\begin{equation}
\label{eq:unitary tunable}
(u_1\otimes u_2) e^{i a\, \sigma_x\otimes\sigma_x + ib\,\sigma_y\otimes\sigma_y + i c\,\sigma_z\otimes\sigma_z} (u_3\otimes u_4)
\end{equation}
with $u_i \in U(2)$ and coefficients $a,b,c\in\mathbb{R}$. $\sigma_i$ denote the Pauli matrices. We define a probability distribution for all $U_i,V_i\in U(4)$ composing the time evolution operator by means of this form, drawing each $u_i$ from the Haar measure for $U(2)$ and $a,b,c$ uniformly from the interval $[-h,h]$. Note that this distribution possesses single-site Haar invariance, so that the results of section~\ref{sec: results spins analytical} still apply.

\begin{figure}
 \centering
 \includegraphics[width=\columnwidth]{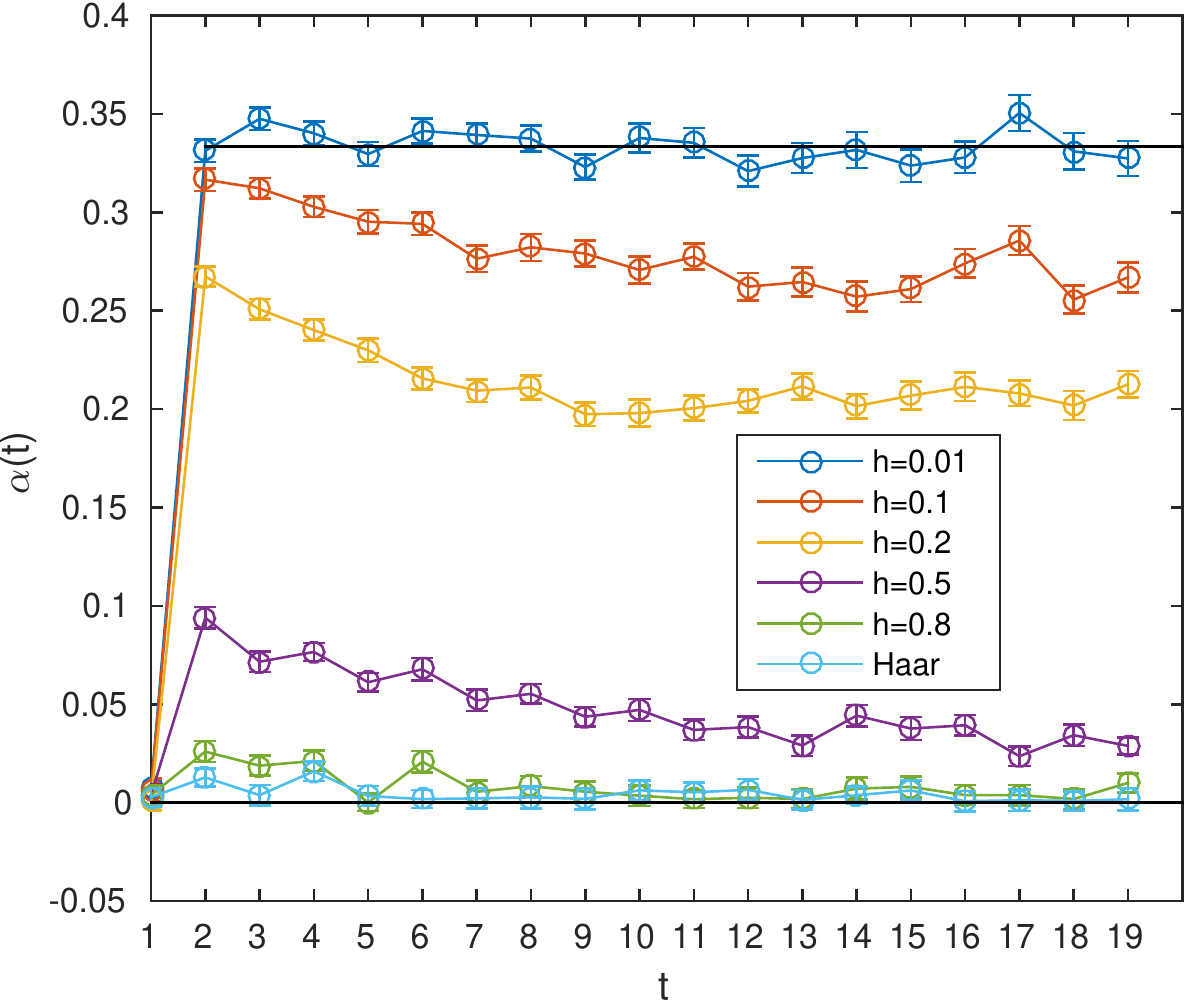}
 \caption{The constant $\alpha(t)$ relating initial and final states (\ref{eq:unitary depolarising channel}) for unitaries distributed according to (\ref{eq:unitary tunable}) with random coupling strength $h$. For strong coupling $\alpha(t)$ relaxes to zero and the system thermalizes. In contrast $\alpha(t)$ reaches a finite value at weak couplings and the system displays localization. Horizontal lines indicate $\alpha= 0$ and the exact decoupled value $h=0,\alpha = 1/3$. See section~\ref{sec:unitary details numeric} for the numerical method used.}
 \label{fig:unitary alphas}
\end{figure}

In Fig.~\ref{fig:unitary alphas} we present numerical results for $\alpha(t)$ for distributions with various coupling strengths $h$.  In the figure, we find a crossover from thermalization for large coupling where $\alpha(t)\to0$ and localization for small coupling where $\alpha(t)$ reaches a finite value and the map~(\ref{eq:unitary depolarising channel}) keeps information about the initial state. In the completely uncoupled case $h=0$, $\alpha = 1/3$ is reached exactly (appendix~\ref{sec:unitary appendix uncoupled}), consistent with the behaviour for $h\to0$.

\begin{figure}
 \centering
 \includegraphics[width=\columnwidth]{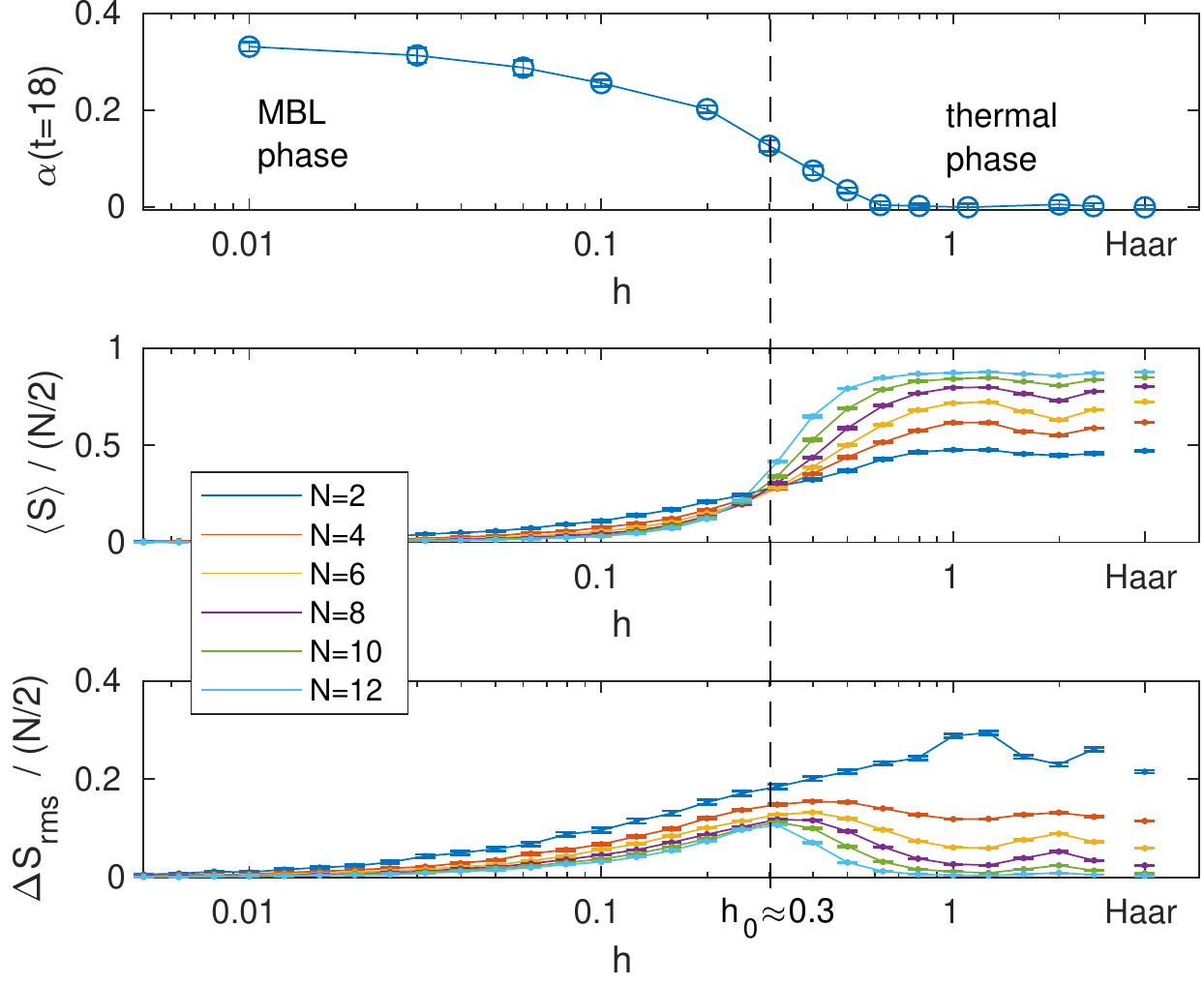}
 \caption{MBL transition at $h_0\approx0.3$.\\%
 Top plot: Late-time value $\alpha(t=18)$ (as in Fig.~\ref{fig:unitary alphas}) as a function of the random coupling strength $h$, and (rightmost da\-ta\-points) for Haar-distributed unitaries.\\%
 Middle and bottom plots:  Average bipartite entanglement entropy (base 2) of eigenstates of $10^3$ samples of the time evolution operator $U$, for several system sizes $N$. In addition to the average entropy of all eigenstates of a random instance $U$ (middle) we calculate the standard deviation of the eigenstates of an instance (bottom). These measures show clear signals of an MBL transition that become more pronounced as the chain length $N$ increases. }
 \label{fig:eigenvector entanglement}
\end{figure}

The MBL transition can be extracted from $\alpha(t=\infty)$ as a function of $h$. Alternatively, it may be pin-pointed by considering the entanglement entropy of the time evolution operator's eigenstates in the limit of an infinite system. In the thermal phase, the eigenstates have volume law entanglement while in the MBL phase they have lower area law entanglement \cite{MBL_mobility_edge, Bauer_Nayak}. Results obtained from exact diagonalization of small systems are shown in Fig.~\ref{fig:eigenvector entanglement}  alongside $\alpha(t=18)$.  It is interesting to consider also the variance of the different eigenstates' entanglement, also plotted in Fig.~\ref{fig:eigenvector entanglement}. Because all eigenstates have similar entanglement properties in both thermal and MBL phases, the variance peaks near the phase transition where the entanglement is intermediate between these limits in a way that varies strongly between eigenstates \cite{Pollmann_MBL_entanglement_entropy}.
Those measures all agree and clearly indicate a finite-size or finite-time estimate of the MBL transition at coupling strength near $h_0\approx 0.3$. Such estimates are known to drift systematically towards the MBL phase as the size of the system is increased \cite{Huse_Floquet}, as can be seen from the crossings in the middle panel of Fig.~\ref{fig:eigenvector entanglement}, so the actual phase transition is most likely at a value of $h$ smaller than this.

\section{Proofs}
\label{sec:proofs}

In this section, we give detailed proofs for the analytic results reported above. The numerical method is explained in the section after.
First, we present a technique used throughout that we call the twirling technique (section~\ref{sec:twirling technique}).
Then we show our results for Gaussian circuits, under inhomogeneous evolution in section~\ref{sec:fermions details inhomogeneous} and homogeneous evolution in section~\ref{sec:fermions details homogeneous}. In the latter case, we also explain the complementary viewpoint provided by eigenvector delocalization. Finally, we proof the results for spin chains in section~\ref{sec:unitary details analytic}.

\subsection{Twirling technique}
\label{sec:twirling technique}

\begin{figure}
 \centering
 \includegraphics[width=\columnwidth]{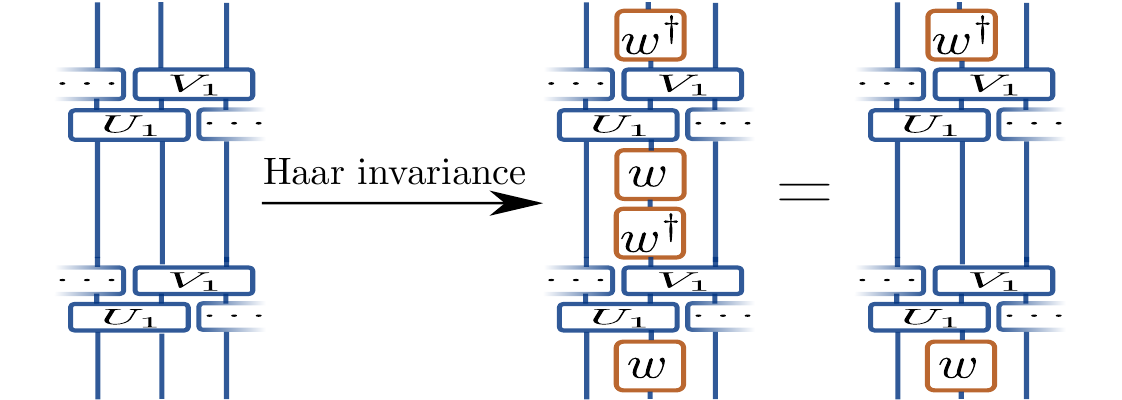}
 \caption{Illustration of the twirling technique (see appendix~\ref{sec:twirling technique}). By single-site Haar invariance, replacing $U_1\to (\Id_d\otimes w)U_1$ and $V_1\to V_1(w^\dagger\otimes\Id_d)$, $w\in U(d)$ does not affect the averaged result. The sketch shows an excerpt of the time evolution operator $U$. In a repeated application $U^t$, most $w$ and $w^\dagger$ cancel.}
 \label{fig:unitary singlesitehaar}
\end{figure}

In this section, we present a technique we call twirling technique, which recurs in the proofs of our results.
The idea is to exploit single-site Haar invariance of the probability distribution for the unitaries.
Single-site Haar invariance means that any transformation of a $U_i$ or $V_i$ of the form  \begin{equation}
U_i \leftrightarrow (w_1\otimes w_2)U_i(w_3\otimes w_4)
\end{equation}
does not affect averages $\avg{\cdot}$, for arbitrary choice of $w_j\in U(d)$.

Our procedure is depicted in Fig.~\ref{fig:unitary singlesitehaar}. At any site $2n$ (here we demonstrate for even sites), we perform the transformation
\begin{equation}
U_{n} \to (\Id_d\otimes w_{2n})U_{n}; V_{n} \to V_{n}(w_{2n}^\dagger\otimes\Id_d)
\end{equation}
with arbitrary $w_{2n}\in U(d)$.
Then $w_{2n}$ cancels with $w_{2n}^\dagger$ in a repeated application of the time evolution operator $U$, which transforms as
\begin{equation}
 U^t \to w_{2n}^\dagger U^t w_{2n}, \label{eq:twirling technique time evolution trafo}
\end{equation}
$w_{2n}$ only acting on site $2n$.

Thus Haar invariance allows us to relate the initial state to the average final state:
\begin{align}
 \avg{\rho_t} &= \avg{U^t\rho_0U^{t\dagger}} = \avg{w_{2n}^\dagger U^tw_{2n}\rho_0 w_{2n}^\dagger U^{t\dagger}w_{2n}} \label{eq:main idea}\\
 &= w_{2n}^\dagger\avg{\rho'_t}w_{2n}\ \text{with}\ \rho_0' = w_{2n} \rho_0 w_{2n}^\dagger.
\end{align}
This holds for arbitrary $w_{2n}\in U(d)$ and can be iterated independently at each site. In some cases, it will prove useful to integrate over $w_{2n}$ in (\ref{eq:main idea}), which, again, does not alter the result $\avg{\cdot}$. An important simplification arises when tracing over sites of the final state, because then in (\ref{eq:main idea}) the left- and rightmost $w_{2n}^\dagger$ and $w_{2n}$ cancel.
In this paper we consider only distributions with single-site Haar invariance. Even in its absence, for example if transformation only with certain $w_{2n}$ are allowed, some results may carry over.

\subsection{Gaussian circuits: Inhomogeneous evolution}
\label{sec:fermions details inhomogeneous}

In this section, we show the result (\ref{eq:fermion inhomogeneous result}). First, we take $P_i,Q_i\in O(4)$. Then, we show how to reduce $P_i,Q_i \in SO(4)$ to the former case.

\subsubsection{Haar measure on orthogonal group}
By linearity of time evolution and Haar-averaging, it suffices to consider only initial states $\Gamma_0^n = \bigoplus_{i=1}^{N} \delta_{in} \gamma$ having all but site $n$ maximally mixed. The $2\times2$ covariance matrix for the occupied site is given by $\gamma = \left(\begin{smallmatrix}0&1\\-1 & 0\end{smallmatrix}\right)$.

We adapt the twirling technique (section~\ref{sec:twirling technique}) to the setting of Gaussian circuits to show that all components of $\avg{\Gamma_t^n}$ are zero, except the $2\times 2$ block corresponding to on-site correlations at site $n$.
To this end, consider the transformation
\begin{equation}
 \bigoplus P_i \to \left(\bigoplus P_i\right) \Sigma;\ \bigoplus Q_i \to G^\dagger\Sigma G\left(\bigoplus Q_i\right)
\end{equation}
with a diagonal matrix $\Sigma$ of signs $\pm 1$. Because $\Sigma$ and $G^\dagger\Sigma G$ have the correct structure to be split among the $P_i$ and $Q_i$, in spirit of the twirling technique we may perform this transformation using the Haar invariance of the Haar-distributed $P_i,Q_i$. Specifically, fix $\Sigma_{2n-1,2n-1} = \Sigma_{2n,2n} = +1$ such that $\Sigma \Gamma_0^n \Sigma = \Gamma_0^n$. Then, similarly to (\ref{eq:main idea}), single-site Haar invariance implies that
\begin{equation}
 \avg{\Gamma_t^n} = \avg{\Sigma O^t\Sigma\Gamma_0^n\Sigma O^{t\dagger}\Sigma} = \Sigma\avg{\Gamma_t^n}\Sigma.
\end{equation}
For each $i\neq 2n-1,2n$, we are free to choose $\Sigma_{i,i} = -1$ and all other signs positive. From this we learn that the entire $i$'th row and $i$'th column (except the diagonal entry) of $\avg{\Gamma_t^n}$ are zero. The only matrix elements that can be non-zero are the diagonal and the $2\times2$ block corresponding to site $n$.

Moreover, the final covariance matrix is real antisymmetric, so the diagonal is also zero and only two entries $\avg{\Gamma_t^n}_{2n-1,2n},\avg{\Gamma_t^n}_{2n,2n-1}$ remain. These form an antisymmetric $2\times2$ block at site $n$. Therefore this block is proportional to the same block of the initial covariance matrix; we can write
\begin{equation}
 \avg{\Gamma_t^n} = c(t,N,n)\Gamma_0^n.
\end{equation}

It remains to show that $c(t,N,n)$ are equal for all $n$. The Haar average treats all unitaries on equal footing, such that within an average $O$ possesses translational invariance by two sites. Therefore $\avg{\Gamma^{n+2}_t} = c(t,N,n)\avg{\Gamma^{n+2}_0}$, mandating that there can only be two distinct values for $n$ even or odd.

Inversion of the chain corresponds to 
\begin{equation}
  P_i \to \left(\begin{smallmatrix} & \Id_2 \\ \Id_2 & \end{smallmatrix}\right)  P_{N/2-i} \left(\begin{smallmatrix} & \Id_2 \\ \Id_2 & \end{smallmatrix}\right),
\end{equation}
and accordingly for $Q_i$. It is a symmetry because in the average, $\left(\begin{smallmatrix} & \Id_2 \\ \Id_2 & \end{smallmatrix}\right)$ can be Haar-absorbed by $P_{N/2-i}$. Inversion invariance implies that there is only one value $c(t,N) = c(t,N,n)$ for both $n$ even and odd, the required form for (\ref{eq:fermion inhomogeneous result}). Any uncorrelated initial state can be decomposed as a linear combination of $\Gamma_0^n$'s, so by linearity, (\ref{eq:fermion inhomogeneous result}) holds with the same constant $c(t,N)$ for each initial state.

We state a few further conclusions.
Note that the particle number of the state changes. Precisely we can formulate
\begin{equation}
 c(t,N) = \frac{n(\avg{\Gamma_t})/N - 1/2}{n(\Gamma_0)/N -1/2}
\end{equation}
because the particle number $n(\Gamma) = \sum (\lambda_i/2+1/2) $ is related to the sum of every second entry $\lambda_i$ along the state's first offdiagonal. Although we consider Gaussian circuits, transformations beyond those preserving particle number are of paramount importance in this setting (as $c(t,N)\neq1$) and our results go beyond mere Anderson localization of a single particle.

After one time step, $c(t=1,N)= 0$ exactly, as direct integration of $P_i,Q_i$ shows \cite{Weingarten_orthogonal}. After that, it becomes non-zero. To assess the localising properties of the system, the long-time behaviour of $c(t,N)$ is of interest. For this, its time average  \begin{equation}
c(\tavg,N) := \lim_{T\to\infty} \frac{1}{T} \sum_{t=0}^{T-1} c(t,N)
\end{equation}
is a useful value. This removes the dependence on the eigenvalues $e^{i\theta_i}$ of $O$: From (\ref{eq:fermion inhomogeneous result}) we can write 
\begin{equation}
 c(t,N) = -\frac{1}{2} \avg{\Tr{\Gamma_0^n O^t\Gamma_0^nO^{t\dagger}}}
 \label{eq:fermion appendix before spectral decomposition}
\end{equation}
for any $1\leq n\leq N$.
Inserting the spectral decomposition $O=\sum_i\ket{v_i} e^{i\theta_i}\! \bra{v_i}$, the generic non-degenerate case $\theta_i\neq\theta_j$ yields
\begin{align}
 \label{eq:fermion time average}
 c(\tavg,N) &= -\frac{1}{2}\bigg\langle\sum_{i,j} \overbrace{\lim_{T\to\infty}\frac{1}{T}\sum_{t=0}^{T-1}e^{i\theta_it-i\theta_jt}}^{\delta_{ij}} \\
 & \phantom{=} \Tr \Gamma_0^n\dyad{v_i}{v_i}\Gamma_0^n\dyad{v_j}{v_j} \bigg\rangle \nonumber\\
 & = \frac{1}{2}\left\avg{ \sum_{i=1}^{2N} | \braket{v_i|\Gamma_0^n|v_i} |^2 \right} \geq 0
\end{align}
While it is expected that this average is strictly positive, its scaling in the thermodynamic limit $N\to\infty$ is unclear.
To establish that $c(t,N)$ reaches a finite value and localization holds, we resort to numerical calculations of $c(t,N)$ as shown in Fig.~\ref{fig:fermion c(t,N)}.

\subsubsection{Haar measure on special orthogonal group}
We will now show that the results for the $O(4)$ Haar measure equally apply when using the $SO(4)$ Haar measure.
Let the number of timesteps $t<(N-1)/2$, such that the lightcone fits into the periodic system without overlapping. Relate the orthogonal to the special orthogonal group by writing $P_i,Q_i\in O(4)$ in the form
\begin{equation}
 P_i = \left(\begin{smallmatrix}1&&& \\ & 1 && \\ & & 1& \\ &&&-1\end{smallmatrix}\right)^{p_i}\tilde{P}_i,\  Q_i = \tilde{Q}_i\left(\begin{smallmatrix}1&&& \\ & -1 && \\ & & 1& \\ &&&1\end{smallmatrix}\right)^{q_i} 
\end{equation}
with $\tilde{P}_i,\tilde{Q}_i\in SO(4)$ and $p_i,q_i\in \mathbb{Z}_2$. Note that the orthogonal Haar distribution for $P_i$ corresponds to the special orthogonal Haar distribution for $\tilde{P}_i$ in combination with the uniform distribution for $p_i$.

Our strategy will consist in showing that the average
\begin{gather}
 \avg{O^t \Gamma_0^n O^{t\dagger} }_{P_i,Q_i \in O(4)} = %
 \left\langle\avg{O^t \Gamma_0^n O^{t\dagger} }_{\tilde{P}_i,\tilde{Q}_i \in SO(4)}\right\rangle_{p_i,q_i\in\mathbb{Z}_2} \label{eq:so=0 pq avg}\\
 = \avg{O^t \Gamma_0^n O^{t\dagger} }_{\substack{\tilde{P}_i,\tilde{Q}_i \in SO(4)\\ p_i,q_i\ \text{fixed}}} = \avg{O^t \Gamma_0^n O^{t\dagger} }_{P_i,Q_i \in SO(4)} \label{eq:so=o pq fix}
\end{gather}
is independent of how $p_i, q_i$ are fixed, i.e.~the equality of (\ref{eq:so=0 pq avg}) and (\ref{eq:so=o pq fix}). Then the equality of $O(4)$ and $SO(4)$ averages immediately follows for all states; these can can be written as linear combinations of $\Gamma_0^n$'s. We may absorb all $q_i$ into $p_i$ by the transformation $q_i\to 0, p_i \to p_i+q_i$ which uses associativity of matrix multiplication to regroup $q_i$ from $\tilde{Q}_i$ to $p_i$ and $\tilde{P}_i$.

Thanks to $SO(4)$ Haar invariance of $\tilde{Q}_k$ and $\tilde{P}_{k+1}$ (for each index $k$ in turn), whenever $p_k=1$ we may perform the transformation $p_k\to0, p_{k+1} \to p_{k+1} + 1$. Specifically, this follows from the $SO(4)$-Haar-invariant transformations
\begin{equation}
 \tilde{Q}_k \to \tilde{Q}_k\left(\begin{smallmatrix}1&&&\\ &-1&& \\ &&1& \\ &&&-1  \end{smallmatrix}\right), \tilde{P}_{k+1} \to \left(\begin{smallmatrix}1&&&\\ &-1&& \\ &&1& \\ &&&-1  \end{smallmatrix}\right) \tilde{P}_{k+1}.
\end{equation}
Iterating this transformation for increasing values of $k$, we may set all $p_i = 0$ except for $p_{N/2}$ which may be $0$ or $1$.
Due to the lightcone size, all occurences of $p_{N/2}$ are multiplied by the zeros in the initial state $\Gamma_0^n = \bigoplus_i \delta_{in}\gamma$ (for $n=N/2$, other initial sites $n$ follow similarly). In conclusion, the average (\ref{eq:so=o pq fix}) is independent of all $p_i,q_i$ and the main result (\ref{eq:fermion inhomogeneous result}) holds for both the $O(4)$ and $SO(4)$ Haar measures.

\subsection{Gaussian circuits: Homogeneous evolution}
\label{sec:fermions details homogeneous}

In this section, the time evolution operator $O$ is homogeneous, $P = P_i = P_j\in O(4), Q = Q_i = Q_j\in O(4)$. We show the result (\ref{eq:fermions homogeneous result}) summarized in section~\ref{sec:fermions results homogeneous}.

\paragraph{Fourier transformation of problem.}
First, let us perform a Fourier transformation of the problem. 
The periodic structure of $O$ suggests a Fourier transform of two-site blocks with $\mathcal{F}\otimes\Id_4$, employing the $N/2\times N/2$ discrete Fourier matrix
\begin{equation}
 \mathcal{F}_{kj} := \frac{1}{\sqrt{N/2}} \exp\left(-2\pi i \frac{(k-1)(j-1)}{N/2}\right),
\end{equation}
where $k,j=1\ldots N/2$. We will denote Fourier transformed quantities with a hat.
The time evolution operator $O$ is block-circulant (\ref{eq:fermion evolution operator}), hence its Fourier transform is block-diagonal and can be written in terms of the diagonal components
\begin{equation}
 O \xrightarrow{\mathcal{F}\otimes\Id_4} \bigoplus_{k=1}^{N/2} \hat{O}_k,\ \hat{O}_k = \hat{G}_k Q\hat{G}_k^\dagger P \in U(4),
 \label{eq:fermion homogeneous Ok}
\end{equation}
with $\hat{G}_k := \left(\begin{smallmatrix} 0& \exp(2\pi i k/(N/2))\Id_2 \\ \Id_2 & 0 \end{smallmatrix} \right)$.

\paragraph{Localized initial state.}
Consider now the localized initial state $\Gamma_0^n = \bigoplus_{i=1}^N \delta_{in}\gamma$ with site $n$ occupied (wlog we consider $n$ odd) and all other sites maximally mixed.
Its Fourier transform is not block-diagonal as for the time evolution operator (\ref{eq:fermion homogeneous Ok}), but also has off-diagonal blocks
\begin{equation}
 \Gamma_0^n \xrightarrow{\mathcal{F}\otimes\Id_4}  (\hat{\Gamma}_0^n)_{kl} = \frac{1}{N/2}e^{i\phi_{kl}} \gammazero
\end{equation}
with phases $\phi_{kl} = 2\pi (n-1)(k-l) /N$.
Accordingly, the final state has off-diagonal blocks, too:
\begin{equation}
 \avg{\Gamma_t^n} \xrightarrow{\mathcal{F}\otimes\Id_4} \avg{\hat{\Gamma}_t^n}_{kl} = \frac{1}{N/2} e^{i\phi_{kl}} \avg{\hat{O}_k^t \gammazero \hat{O}_l^{t\dagger}}.
 \label{eq:fermion homogeneous gamma_kl}
\end{equation}
Numerical calculations provide evidence that all blocks $\avg{\hat{\Gamma}_t^n}_{kl}$ vanish as $t\to\infty$, except for the diagonal $k=l$, and the pairs $(k,l) = (N/2, N/4), (k,l) = (N/4,N/2)$. These pairs only exist for $N$ divisible by four and correspond to the Fourier phases $0$ and $\pi$. Only for these two pairs are both $\hat{O}_k$ and $\hat{O}_l$ real.

\paragraph{Time-average for localized initial state.}
As we now argue, in the time average $\avg{\Gamma_\tavg^n}$ all Fourier blocks $\avg{\hat{\Gamma}_\tavg^n}_{kl}, k\neq l, (k,l)\neq (N/2,N/4),(k,l)\neq(N/4,N/2)$ are zero. 
Inserting the spectral decomposition $\hat{O}_k = \sum_{i=1}^4 e^{i\theta_{k,i}}\dyad{v_{k,i}}{v_{k,i}}$, the time average is
\begin{multline}
 \avg{\hat{\Gamma}_\tavg^n}_{kl} = \frac{e^{i\phi_{kl}}}{N/2} \bigg\langle \sum_{i,j=1}^4\overbrace{\lim_{T\to\infty}\frac{1}{T}\sum_{t=0}^{T-1} e^{i(\theta_{k,i} - \theta_{l,j})t}}^{(*)} \\ \dyad{v_{k,i}}{v_{k,i}} \gammazero \dyad{v_{l,j}}{v_{l,j}} \bigg\rangle.
 \label{eq:fermion homogeneous offdiagonal tavg}
\end{multline}
For each fixed pair $(k,l)$, whenever the sets of eigenvalues $\{e^{i\theta_{k,i}},i=1,2,3,4\}$ of $\hat{O}_k$ and $\{e^{i\theta_{l,i}},i=1,2,3,4\}$ of $\hat{O}_l$ are disjoint, $(*)$ is zero. Moreover, for the Haar average $\avg{\hat{\Gamma}^n_\tavg}_{kl}$ to vanish for any given pair $(k,l)$, it suffices that the eigenvalue sets are disjoint for all $P,Q$ except a measure zero set. We conjecture that this holds for all pairs $(k,l), k\neq l$ and $(k,l)\neq(N/2,N/4),(k,l)\neq(N/4,N/2)$ \footnote{%
It is interesting to understand why the statement does not hold for $k=l$ and the two specific pairs. For $k=l$, it is obvious that $\hat{O}_k = \hat{O}_l$ are identical matrices and have identical spectra.
The pairs $(k,l) = (N/2,N/4),(N/4,N/2)$ are the only values for which both $\hat{G}_k$ and $\hat{G}_l$ are real. As real orthogonal matrices, the eigenvalues of $\hat{O}_{k/l}$ are real ($\pm1$) or arise as complex-conjugate pairs. In fact, for a quarter of all choices $(P,Q)$, the determinants are $\det\hat{O}_{k} = \det\hat{O}_l = -1$. These determinants force both matrices to both have an eigenvalue $+1$ and $-1$. 
}.

In conclusion, in the time average of (\ref{eq:fermion homogeneous gamma_kl}) of an initially localized state, only Fourier components $k=l$ and $k,l=N/2,N/4$ survive:
\begin{subequations}
\begin{align}
 \avg{\Gamma_\tavg^n} & \xrightarrow{\mathcal{F}\otimes\Id_4} \avg{\hat{\Gamma}_\tavg^n}_{kl} = \\
 & \frac{\delta_{k=l}}{N/2}\lim_{T\to\infty}\frac{1}{T}\sum_{t=0}^{T-1}\avg{\hat{O}_k^t\gammazero \hat{O}_k^{t\dagger}}  \label{eq:Gtavgn ti part} \\
 &+ \frac{\delta_{k=N/2,l=N/4}+\delta_{k=N/4,l=N/2}}{N/2} \\
 &\quad e^{i\phi_{kl}} \lim_{T\to\infty}\frac{1}{T}\sum_{t=0}^{T-1}  \avg{\hat{O}_k^t\gammazero \hat{O}_l^{t\dagger}}. \label{eq:Gtavgn 0pi part2}
\end{align}
\end{subequations}
Note that the position $n$ of the initial localization is present in (\ref{eq:Gtavgn ti part}) only as $n\!\!\mod2$, determining $\gammazero$ or $\zerogamma$, and in (\ref{eq:Gtavgn 0pi part2}) only as $n\!\!\mod4$, in the phase $\phi_{kl} = 2\pi(n-1)(k-l)/N = \pm \pi/2\,(n-1)$.

The Fourier backtransformation is
\begin{equation}
 \avg{\Gamma_\tavg^n} = \frac{1}{N/2} \Gamma_\star := \frac{1}{N/2}(\Gamma_\star' + \Gamma_\star''),
\end{equation}
where $\Gamma_\star'$ denotes the backtransform arising from the diagonal Fourier components (\ref{eq:Gtavgn ti part}) and $\Gamma_\star''$ the backtransfom arising from the Fourier components $k,l=N/2,N/4$  (\ref{eq:Gtavgn 0pi part2}), each without the prefactor $1/(N/2)$.
As discussed in the previous paragraph, $\Gamma_\star$ is only dependent on $n\!\!\mod4$ and, thanks to the prefactor, vanishes in the limit $N\to\infty$. This concludes our result (\ref{eq:fermions homogeneous result}).

\paragraph{Translation-invariant initial state.}
On top of the results summarized in section~\ref{sec:fermions results homogeneous}, we provide a characterization of $\Gamma'_\star$. For this, we need to first consider the trans\-la\-tion-in\-vari\-ant initial state $\Gamma_0^\text{t.i.}$ with each site occupied. It is invariant under Fourier transformation:
\begin{equation}
 \Gamma_0^\text{t.i.} = \bigoplus_{i=1}^{N} \gamma \xrightarrow{\mathcal{F}\otimes\Id_4} \hat{\Gamma}_0^\text{t.i.} = \bigoplus_{k=1}^{N/2} \gammagamma.
\end{equation}
The final state has a block-diagonal Fourier transform and each block has the form
\begin{equation}
  \avg{\Gamma_t^\text{t.i.}} = \avg{O^t \Gamma_0^\text{t.i.} O^{t\dagger}}\xrightarrow{\mathcal{F}\otimes\Id_4} \avg{\hat{\Gamma}_t^\text{t.i.}}_k = \avg{ \hat{O}_k^t \gammagamma \hat{O}_k^{t\dagger} }.
  \label{eq:evolution periodic initial}
\end{equation}
We will now relate the translation-invariant state's time average $\avg{\Gamma_\tavg^\text{t.i.}}$ to $\Gamma_\star'$ in the time average of localized initial states.

\paragraph{Relating $\Gamma_\star'$ to $\avg{\Gamma_\tavg^\textup{t.i.}}$.}
We now show that $\Gamma_\star' = \avg{\Gamma_\tavg^\text{t.i.}}$ except that all rows and columns corresponding to even sites are zero. For this, we exploit Haar invariance, similarly to the twirling technique and the case of inhomogeneous fermionic time evolution. The twirling technique can be used either with the original time evolution operator (\ref{eq:fermion evolution operator}) and (\ref{eq:fermion final state def}) or equivalently directly in the Fourier transformed quantities (\ref{eq:fermion homogeneous Ok}) and (\ref{eq:evolution periodic initial}), which in the following is the perspective we take.

First, we show that $\avg{\hat{\Gamma}_t^\text{t.i.}}_k$ (see (\ref{eq:evolution periodic initial})) is $2\times2$ block-diagonal. This follows from the transformation
\begin{equation}
 P \to P\Sigma, Q \to \hat{G}_k^\dagger\Sigma^\dagger\hat{G}_k Q
 \label{eq:fermion details homogeneous trafo}
\end{equation}
with $\Sigma = \left(\begin{smallmatrix}\pm\Id_2 & \\ & \pm\Id_2\end{smallmatrix}\right)$. Note that $\hat{G}_k^\dagger\Sigma^\dagger\hat{G}_k$ is real orthogonal as required to apply $O(4)$ Haar invariance.
This transformation effects, as in (\ref{eq:main idea}),
\begin{equation}
 \avg{\hat{\Gamma}_t^\text{t.i.}}_k = \avg{\Sigma^\dagger \hat{O}_k^t\Sigma \gammagamma \Sigma^\dagger \hat{O}_k^{t\dagger}\Sigma} = \Sigma^\dagger \avg{\hat{\Gamma}_t^\text{t.i.}}_k \Sigma.
 \label{eq:fermion details homogeneous trafo effect}
\end{equation}
With appropriate choice of signs in $\Sigma$ it follows that the off-diagonal blocks of $\avg{\hat{\Gamma}_t^\text{t.i.}}_k$ vanish.

Now we are in a position to show the relation between $\avg{\hat{O}_k^t\gammagamma \hat{O}_k^{t\dagger}}$ and $\avg{\hat{O}_k^t\gammazero \hat{O}_k^{t\dagger}}$ appearing in the Fourier transformations of $\avg{\Gamma_\tavg^\text{t.i.}}$ and $\Gamma_\star'$, respectively.
For this, use the transformation (\ref{eq:fermion details homogeneous trafo}) with
\begin{equation}
 \Sigma = \left(\begin{smallmatrix}1 & & & \\ & 1 & & \\ & &0 & 1 \\ & &1 &0\end{smallmatrix}\right).
 \label{eq:fermion details homogeneous trafo sigmax}
\end{equation}
Again, note $\hat{G}_k^\dagger\Sigma\hat{G}_k$ is real. Then
\begin{align}
 &\avg{\hat{O}_k^t\gammazero \hat{O}_k^{t\dagger}}
 = \frac{1}{2} \left[\avg{\hat{O}_k^t\gammagamma \hat{O}_k^{t\dagger}} +
 \avg{\hat{O}_k^t\left(\begin{smallmatrix}\gamma & \\ & -\gamma\end{smallmatrix}\right) \hat{O}_k^{t\dagger}} \right] \\
 &= \frac{1}{2} \left[\avg{\hat{O}_k^t\gammagamma \hat{O}_k^{t\dagger}} +
 \avg{\hat{O}_k^t \Sigma\gammagamma\Sigma \hat{O}_k^{t\dagger}} \right] \nonumber\\
 &= \frac{1}{2} \left[\avg{\hat{O}_k^t\gammagamma \hat{O}_k^{t\dagger}} +
 \Sigma \avg{\hat{O}_k^t \gammagamma \hat{O}_k^{t\dagger}} \Sigma \right],  \nonumber
\end{align}
with the last equality due to Haar invariance.
Thanks to one term without and one term conjugated by $\Sigma$, the first block stays the same and the second block (which corresponds to even sites) cancels. This carries through the Fourier transform $\mathcal{F}\otimes\Id_4$, concluding our proof that $\Gamma_\star' = \avg{\Gamma_\tavg^\text{t.i.}}$ except that all rows and columns corresponding to even sites are zero.

\paragraph{Characterization of $\avg{\Gamma_t^\text{t.i.}}$.}
Now we will further characterize $\avg{\Gamma_t^\text{t.i.}}$ arising from the translation-invariant initial state.
In terms of the Fourier components of $\avg{\Gamma_t^\text{t.i.}}$, we show below that
\begin{equation}
 \avg{\hat{\Gamma}_t^\text{t.i.}}_k = c(t,k,N)\gammagamma,
 \label{eq:fermion homogeneous characterization Gamma result}
\end{equation}
with a real constant $c(t,k,N)$.
This is a similar form as the main result (\ref{eq:fermion inhomogeneous result}) except that there is one constant per Fourier component. The methods in the proof following are also very similar.

We have shown already that $\avg{\hat{\Gamma}_t^\text{t.i.}}_k$ consists of two $2\times2$ blocks, in the paragraph of equation~(\ref{eq:fermion details homogeneous trafo}).
$\avg{\hat{\Gamma}_t^\text{t.i.}}_k$ is evidently anti-hermitian as a real antisymmetric matrix conjugated with a unitary (\ref{eq:evolution periodic initial}). To prove (\ref{eq:fermion homogeneous characterization Gamma result}), it remains to show that both blocks are real and identical.

First, we show that both blocks are real. To this end, use the transformation (\ref{eq:fermion details homogeneous trafo}) with
\begin{equation}
\Sigma = \left(\begin{smallmatrix}0&1&&\\-1&0&&\\ &&0 &1 \\ && -1 & 0\end{smallmatrix}\right).
\end{equation}
Note that $\hat{G}_k^\dagger\Sigma^\dagger\hat{G}_k$ is real orthogonal, so Haar invariance of $P$ and $Q$'s probability distribution is applicable.  For anti-hermitian $2\times2$ matrices $X$,
\begin{equation}
X = \left(\begin{smallmatrix}0&1\\-1&0\end{smallmatrix}\right)^\dagger X \left(\begin{smallmatrix}0&1\\-1&0\end{smallmatrix}\right) \Rightarrow X\ \text{real antisymmetric}.
\end{equation}
Therefore, the transformation's effect (\ref{eq:fermion details homogeneous trafo effect}) shows that both blocks are real.

Lastly, we show that both blocks are identical. This is achieved by considering inversion symmetry of the chain. Inversion corresponds to
\newcommand{\invmat}{\left(\begin{smallmatrix}& \Id_2 \\ \Id_2 & \end{smallmatrix}\right)}
\begin{equation}
 P \to \invmat P \invmat
\end{equation}
and likewise for $Q$. This is equivalent to
\begin{equation}
 \hat{O}_k \to \invmat \hat{O}_k^* \invmat
 \label{eq:fermion Ok complex conjugation}
\end{equation}
and
\begin{equation}
\avg{\hat{\Gamma}_t^\text{t.i.}}_k \to \invmat \avg{\hat{\Gamma}_t^\text{t.i.}}^*_k \invmat.
\end{equation}
Thus inversion invariance mandates that both real blocks are the same.

\begin{figure}
 \centering
 \includegraphics[width=\columnwidth]{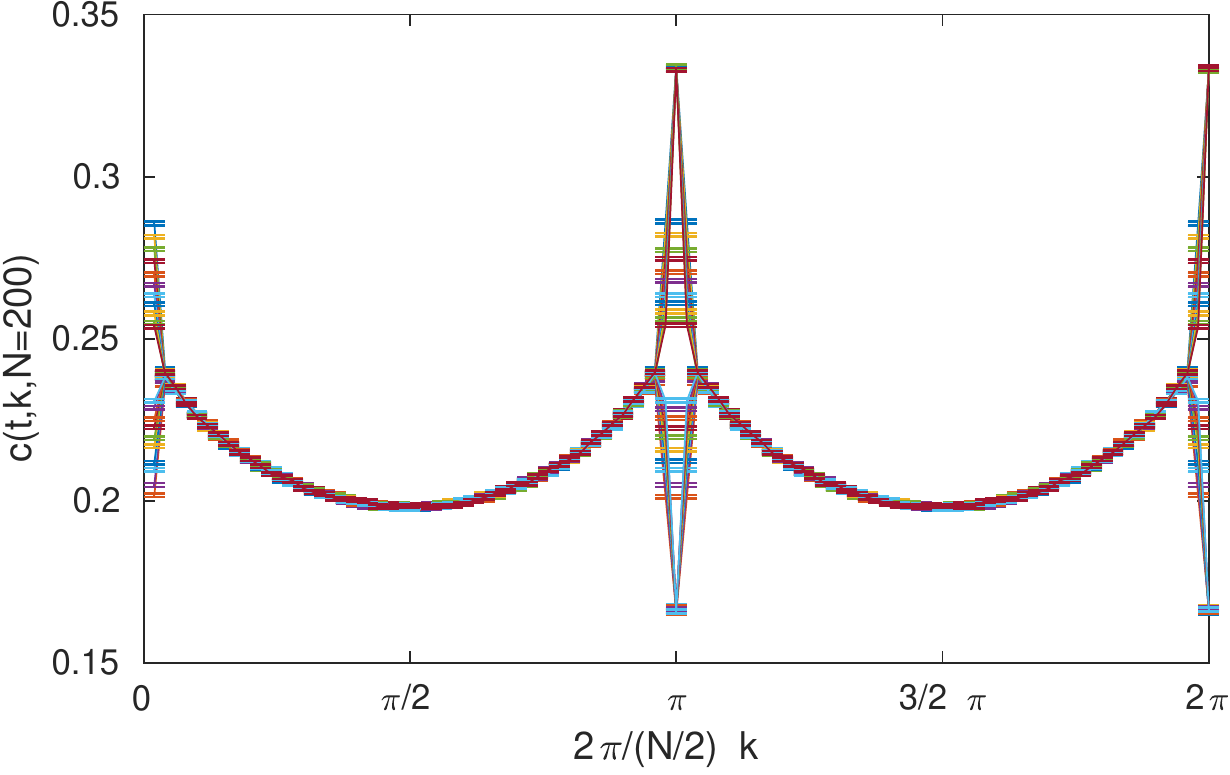}
 \caption{Constant $c(t,k,N=200)$ for the characterization of $\avg{\Gamma^\text{t.i.}_t}$ (\ref{eq:fermion homogeneous characterization Gamma result}) for homogeneous evolution of fermions. Each of the lines shows data for a fixed $t$ from 20 to 50.  We perform the orthogonal Haar average by considering $10^6$ samples. }
 \label{fig:fermion homogeneous c}
\end{figure}

\paragraph{Characterization of $c(t,k,N)$.}
To understand $\avg{\Gamma_t^\text{t.i.}}$, it remains to characterize $c(t,k,N)$.
In Fig.~\ref{fig:fermion homogeneous c} we therefore show numerical calculations of $c(t,k,N)$.

The figure shows that $c(t,k,N)$ converges, except for $\frac{2\pi}{N/2} k = 0,\pi, 2\pi$, where there are oscillations in $t$.
We can calculate the time average of values
\begin{equation}
 c(\tavg,k,N) = \frac{1}{4}\ \text{for}\ \frac{2\pi}{N/2} k = 0,\pi,2\pi
 \label{eq:fermion homogeneous ctkN special values}
\end{equation}
exactly. In these cases, $\hat{G}_k$ is real orthogonal and can be absorbed by the Haar-invariant transformation $Q \to \hat{G}_k^\dagger Q \hat{G}_k$. In turn, $Q$ can be absorbed by the transformation $P \to Q^\dagger P$. Then we have simply $\hat{O}_k = P$; this corresponds to two sites in the uncoupled case and is explained in appendix~\ref{sec:fermion uncoupled}.

The symmetries $k \leftrightarrow -k = N/2 - k$ and ``$k\leftrightarrow k+\pi$'' of $c(t,k,N)$ are apparent in Fig.~\ref{fig:fermion homogeneous c}. The first corresponds to complex conjugation of $\hat{O}_k$, and is a symmetry because $\avg{\Gamma_t^\text{t.i.}}_k$ is real. The latter symmetry only exists for even $N/2$ and then reads $k\leftrightarrow k+N/4$. It is equivalent to the Haar-invariant transformation
\begin{equation}
Q\to \left(\begin{smallmatrix}-\Id_2 & \\ & \Id_2 \end{smallmatrix}\right) Q \left(\begin{smallmatrix}\Id_2 & \\ & -\Id_2 \end{smallmatrix}\right)
\end{equation}
which effects $e^{\pm2\pi i k/(N/2)} \to - e^{\pm2\pi i k/(N/2)}$ in (\ref{eq:fermion homogeneous Ok}).

\subsubsection{Homogeneous evolution --- Eigenvector delocalization}

\label{sec:fermions eigenvector delocalization}

In this section, we explain in detail a complementary viewpoint to delocalization summarized in section~\ref{sec:fermions results homogeneous}, eigenvector delocalization.
The eigenvectors $\vec{v_i}$ to eigenvalues $e^{i\theta_i}$ of each generic instance of the homogeneous evolution operator $O$ are delocalized. To see this, let $\mathcal{T}$ be the orthogonal operator effecting translation by two sites (4 matrix entries). It commutes with $O$ and has $\mathcal{T}^{N/2}=\Id_{2N}$. Each $\vec{v_i}$ is therefore eigenvector of $\mathcal{T}$ to a phase $\phi_i$ and for its components the relation $v_i^{j+4} = e^{\phi_i} v_i^j$ holds circularly. Taking $\vec{v_i}$ normalized, 
\begin{equation}
|v_i^j|^2 = \frac{2}{N} \left(\sum_{k=0}^{N/2-1} |v_i^{j+4k}|^2\right)
\le \frac{2}{N} |\vec{v_i}|^2 = \frac{2}{N}.
\label{eq:fermion homogeneous eigenvector bound}
\end{equation}

In the generic case, where $O$ does not have degenerate eigenvalues, we can give an estimate of the final covariance matrix even without resorting to a Haar average. We expand $\Gamma_\tavg^n$ with the spectral decomposition $O=\sum_k\ket{v_k} e^{i\theta_k}\! \bra{v_k}$ of the specific instance of the time evolution operator:
\begin{align}
 \Gamma^n_\tavg &= \lim_{T\to\infty}\frac{1}{T} \sum_{t=0}^{T-1} O^t \Gamma_0^n O^{t\dagger} \\
  &= \sum_{k,l=1}^{2N} \overbrace{\lim_{T\to\infty}\frac{1}{T}\sum_{t=0}^{T-1} e^{i(\theta_k-\theta_l)t}}^{\delta_{\theta_k,\theta_l}} \dyad{v_k}{v_k}\Gamma_0^n \dyad{v_l}{v_l}.
\end{align}
Similarly to (\ref{eq:fermion time average}), the time average cancels cross terms.

Let $\vec{e_j}$ be the standard basis. The matrix elements of $\Gamma^n_\tavg$ are then
\begin{equation}
 \braket{e_i | \Gamma^n_\tavg | e_j} =\sum_{k=1}^{2N}\braket{e_i|v_k}\braket{v_k|\Gamma_0^n|v_k}\braket{v_k|e_j}.
\end{equation}
With the bound (\ref{eq:fermion homogeneous eigenvector bound}) for the eigenvector's components, we can show the estimate (\ref{eq:fermion results eigenvector delocalization}) by expanding $\Gamma_0^n = \dyad{e_{2n-1}}{e_{2n}} - \dyad{e_{2n}}{e_{2n-1}}$:
\begin{align}
&|(\Gamma^n_\tavg)_{ij}| = |\braket{e_i | \Gamma^n_\tavg | e_j}| \\
&\le \underbrace{\sum_{k=1}^{2N}}_{2N} \underbrace{|\braket{e_i|v_k}|}_{\le \sqrt{2/N}} \underbrace{|\braket{v_k|\Gamma_0^n|v_k}|}_{\le 4/N}\underbrace{|\braket{v_k|e_j}|}_{\le \sqrt{2/N}}
 \le \frac{16}{N}
\end{align}

\subsection{Spins}
\label{sec:unitary details analytic}

In this section we prove our analytic results about spin chains summarized in section~\ref{sec: results spins analytical}, using the twirling technique from section~\ref{sec:twirling technique}. We require the probability distribution for the unitaries $U_i,V_i$ comprising the unitary-circuit time evolution operator to possess single-site Haar invariance, as introduced in the settings~\ref{sec:settings spins}.

The integral over the unitary group
\begin{align}
 \label{eq:haar 2-design}
 & \int_{U(d)}dw_n\, w_n^\dagger A w_n B_n w_n^\dagger C w_n \\
 & = \frac{\Id^n_d}{d} \Tr_n(B_n)\otimes\frac{d\Tr_n(AC) - \Tr_n(A)\Tr_n(C)}{d^2-1} \nonumber\\
 & \phantom{=} + B_n\otimes\frac{d\Tr_n(A)\Tr_n(C) - \Tr_n(AC)}{d(d^2-1)} \nonumber
\end{align}
can be computed exactly \cite{Weingarten_unitary}. Here $A,C\in U(d^N)$ are multi-qudit operators and $w_n,B_n\in U(d)$ act only on one qudit at site $n$. The left side of the tensor products is qudit $n$ while the right side contains all the other sites. The same result is obtained when averaging over a unitary 2-design such as, for qubits, the Clifford group \cite{Clifford_group_3-design} instead of entire $U(d)$.

Similarly, we have the integral
\begin{equation}
 \int_{U(d)}dw_i\, w_i D w_i^\dagger = \Tr_i(D) \otimes \Id^i_d/d,
 \label{eq:haar 1-design}
\end{equation}
where $D\in U(d^N)$ is a multi-qudit operator and $w_i\in U(d)$ acts only on one qudit at site $i$. The identity $\Id^i_d/d$ is at the qudit site $i$ which is traced out from $D$. The integral holds equally for the integrand $w_i^\dagger D w_i$.
For this integral, a unitary 1-design is sufficient for $w_i$, such as for $d=2$ the Pauli matrices together with the identity.

To show our result (\ref{eq:unitary depolarising channel}) for a single-site reduced density matrix at site $n$, let $I = \{1\ldots N\}\backslash\{n\}$ be the set of all other sites. Let $\rho_0$ be the (arbitrary) initial state. With this notation, we will compute the relation between $\Tr_I\rho_0$ and
\begin{equation}
 \Tr_I \avg{\rho_t} = \Tr_I \avg{U^t \rho_0 U^{t\dagger}}.
\end{equation}

At each site $i\in I$ in turn, the twirling technique (\ref{eq:main idea}) results in
\begin{equation}
 \Tr_I \avg{\rho_t} = \Tr_I \avg{ w_i^\dagger U^t w_i \rho_0 w_i^\dagger U^{t\dagger} w_i}
 = \Tr_I \avg{ U^t w_i \rho_0 w_i^\dagger U^{t\dagger} }.
\end{equation}
We may integrate over $w_i$, whose choice is arbitrary, by setting $D=\rho_0$ in formula (\ref{eq:haar 1-design}). This gives
\begin{equation}
 \Tr_I \avg{\rho_t} = \Tr_I \avg{ U^t [\Tr_{i}(\rho_0)\otimes\Id^i_d/d] U^{t\dagger} }.
\end{equation}
Iteration of this procedure for each $i\in I$ yields
\begin{equation}
 \Tr_I \avg{\rho_t} = \Tr_I \avg{ U^t [\Tr_I(\rho_0)\otimes \Id^{I}_{d^{N-1}}/d^{N-1}] U^{t\dagger} }.
\end{equation}

The twirling technique at site $n$ allows us to use formula (\ref{eq:haar 2-design}) with $A = U^t, B_n=\Tr_I(\rho_0), C = U^{t\dagger}$:
\begin{align}
 \Tr_I \avg{\rho_t} &= \frac{1}{d^{N-1}} \Tr_I \avg{ w_n^\dagger U^t w_n \Tr_I(\rho_0) w_n^\dagger U^{t\dagger} w_n } \\
 & = \frac{1}{d^{N-1}} \Tr_I\bigg\langle \frac{\Id^n_d}{d} \Tr_n\!\Tr_I(\rho_0) \,\otimes & \\
 & \phantom{= \frac{1}{d^{N-1}} \Tr_I\bigg\langle} \frac{d\Tr_n(AC) - \Tr_n A \Tr_n C}{d^2-1} \\
 & \phantom{=} + \Tr_I(\rho_0) \otimes \frac{d \Tr_nA\Tr_nC - \Tr_n(AC)}{d(d^2-1)} \bigg\rangle \\
 & =  \frac{\Id_d^n}{d} \frac{ d^2 - \lambda(t)}{d^2-1} + \Tr_I(\rho_0) \frac{\lambda(t) - 1}{d^2-1} \label{eq:unitary details id + reduced} \\
 & = \frac{\Id_d^n}{d} + \underbrace{\frac{\lambda(t)-1}{d^2-1}}_{ \alpha(t) }\bar{\rho}_0^n. \label{eq:unitary details depolarising channel}
\end{align}
In the third equality we have used that $\Tr_n\!\Tr_I(\rho_0) = 1$, $\Tr_I\!\Tr_n(AC) = d^N$ and defined the constant
\begin{equation}
 \lambda(t) = \left\avg{\frac{1}{d^{N-1}} \Tr_I(\Tr_nU^t\Tr_n U^{t\dagger}) \right}
 \label{eq:unitary lambda}
\end{equation}
into which we have moved the remaining Haar average. $\lambda$ is manifestly real and non-negative. In the final equality we have rewritten the expression in terms of the traceless part $\bar{\rho}_0^n$ of the initial reduced density matrix $\Tr_I\rho_0$.
The form (\ref{eq:unitary depolarising channel}) of our result can be obtained by setting
$
 \alpha(t) = \frac{\lambda(t) -1}{d^2 -1}.$

A lightcone structure emerges in the definition of $\lambda$. Only constituent unitaries of $U$ within a lightcone of velocity 2 around site $n$ contribute to $\lambda$, all others cancel with their daggered counterpart in consequence of $\Tr_I$. A longer chain will have an additional $\Tr_i(\Id_d)$ at each additional site $i$ outside the lightcone, which is precisely cancelled by the higher $N$ in the prefactor. (Fig.~\ref{fig:lightcones} shows a graphical representation of a slightly different quantity but also serves to illustrate this fact.)
In combination with Haar invariance of $U$, within the average $\avg{\cdot}$ that treats all constituent unitaries on equal footing, we realize the following. $\lambda(t)$ is independent of site position $n$ or chain length $N$ as long as the lightcone around $n$ does not intersect a boundary, or, in the case of periodic boundary conditions, itself.

After a single timestep, $\lambda(1) = 1$ exactly such that the evolution results in a locally maximally mixed site (\ref{eq:unitary details depolarising channel}). For longer times, we resort to a numerical method for evaluating $\alpha(t) = \frac{\lambda(t)-1}{d^2-1}$, explained in the next section~\ref{sec:unitary details numeric}.

Next let us calculate the entire final density matrix $\avg{\rho_t}$ for the initial state
\begin{equation}
 \rho_0 = \rho_0^n \otimes \Id^I_{d^{N-1}}/d^{N-1}
\end{equation}
that has all sites maximally mixed apart from site $n$.
The twirling technique and formula (\ref{eq:haar 1-design}) with $D = U^t\rho_0U^{t\dagger}$ can be applied at each site $i\in I$ iteratively:
\begin{align}
\avg{\rho_t} &= \avg{ w_i^\dagger U^t w_i \rho_0 w_i^\dagger U^{t\dagger} w_i } 
 = \avg{w_i^\dagger U^t \rho_0 U^{t\dagger} w_i} \\
 &= \avg{ \Tr_i ( U^t \rho_0 U^{t\dagger} ) } \otimes \Id^i_d/d \\
 &= \avg{ \Tr_I (U^t \rho_0 U^{t\dagger}) } \otimes \Id^I_{d^{N-1}}/d^{N-1} \\
 &= \Tr_I\avg{\rho_t} \otimes \Id^I_{d^{N-1}} / d^{N-1}.
\end{align}
All sites of the final state are maximally mixed except for site $n$, it is related to the initial $\rho_0^n = \Tr_I(\rho_0)$ as per (\ref{eq:unitary details id + reduced}).

Let us turn to the behaviour of two-site reduced density matrices for the not necessarily adjacent sites $n$ and $m$, now $I = \{1\ldots N\}\backslash\{n,m\}$. Assume the initial state's reduced density matrix to be a tensor product and split it
\begin{equation}
 \Tr_I(\rho_0) = (\Id^n_d/d + \bar{\rho}_0^n)\otimes(\Id^m_d/d + \bar{\rho}_0^m)
\end{equation}
into traceful and traceless parts.

To determine the final state $\Tr_I\avg{\rho_t}$, we employ the same method as before. However we will have to use formula (\ref{eq:haar 2-design}) twice, at sites $n$ and $m$, and the resulting $\Tr_n$ and $\Tr_m$ terms couple. A calculation yields the compact result
\begin{align}
\label{eq:unitary appendix two-site result}
 \Tr_I\avg{\rho_t} = 
 &\left(\frac{\Id_d^n}{d} + \frac{\lambda-1}{d^2-1}\bar{\rho}_0^n\right)\otimes\left(\frac{\Id_d^m}{d} + \frac{\lambda-1}{d^2-1}\bar{\rho}_0^m\right) \\
 &+ \frac{\lambda' - \lambda^2}{(d^2-1)^2} \bar{\rho}_0^n\otimes\bar{\rho}_0^m. \nonumber
\end{align}
Here $\lambda = \lambda(t)$ is the same as before in (\ref{eq:unitary lambda}), so the first term is simply an uncorrelated tensor product of the single site result (\ref{eq:unitary details depolarising channel}). The coefficient
\begin{equation}
 \lambda'(t) = \left\avg{\frac{1}{d^{N-2}}\Tr_I(\Tr_{n,m}U^t\Tr_{n,m}U^{t\dagger})\right}
\end{equation}
appearing in the second term is also real and positive. It depends on $|n-m|$ until the sites are far enough apart such that their lightcones do not intersect. (This requires a sufficient system size.) In that case, $\lambda' = \lambda^2$ and the two-site result (\ref{eq:unitary appendix two-site result}) reduces to the single site result (\ref{eq:unitary details depolarising channel}).

Our method to show that the evolution of a single site is a depolarising channel (\ref{eq:unitary details depolarising channel}) may readily be generalized to further time evolutions other than the specific quantum circuit considered here. For this, the time evolution operator must allow for transformations of the form (\ref{eq:twirling technique time evolution trafo}), such that the twirling technique can be applied analogously. One such example was studied in \cite{ZnidaricPRL2011,Gessner_Breuer}, which considered a random non-local Hamiltonian coupling all $N$ spins, whose diagonalising matrix is distributed according to the $U(d^{N})$ Haar measure. In that case, the expression for $\alpha(t)$ can be simplified in terms of the spectral form factor of the Hamiltonian.

\section{Numerical method for spins}
\label{sec:unitary details numeric}

In this section, we present the new numerical method we use for the setting of spin chains.
Obtaining numerical values for $\alpha(t)$ of (\ref{eq:unitary depolarising channel})
is much more difficult than for $c(t,N)$ in the fermionic case, because the Hilbert space grows exponentially while covariance matrices grow only quadratically in system size $N$. In the following, we describe a new numerical method that significantly decreases the complexity from $4^{2t+1}$ to $2^t$ for $t$ timesteps, at effectively infinite system size. For definiteness, we set the local Hilbert space dimension $d=2$ although our numerical method can be adapted to higher spins.

We determine $\alpha(t)$ by preparing an initial state where one site is spin up $\ket{0}\!\bra{0}$ and all other sites are maximally mixed. After applying $U^t$, we project the final reduced density matrix of the one site onto $\ket{0}\!\bra{0}$. According to (\ref{eq:unitary depolarising channel}), this procedure yields
\begin{equation}
(\alpha+1)/2 = \avg{R(U,0)}_U,
\end{equation}
\begin{align}
 \label{eq:unitary rhomboid value}
 &R(U,s) = 
 \Tr\big[(\cdots\otimes\Id_2\otimes(\ket{s}\!\bra{s})\otimes\Id_2\otimes\cdots) \\
 &\phantom{=} U^t (\cdots\otimes\Id_2/2\otimes(\ket{0}\!\bra{0})\otimes\Id_2/2\otimes\cdots) U^{t\dagger}\big]. \nonumber
\end{align}
The average $\avg{\cdot}_U$ refers to averaging the random $U_i,V_i$ composing $U$.
Leaving the final spin $s$ free allows us to use an importance sampling technique. Before explaining this technique, we will show how to evaluate $R(U,s)$ for a given $U,s$ in a way that is significantly more efficient than the naive procedure.

\begin{figure}
 \centering
 \includegraphics[width=8cm]{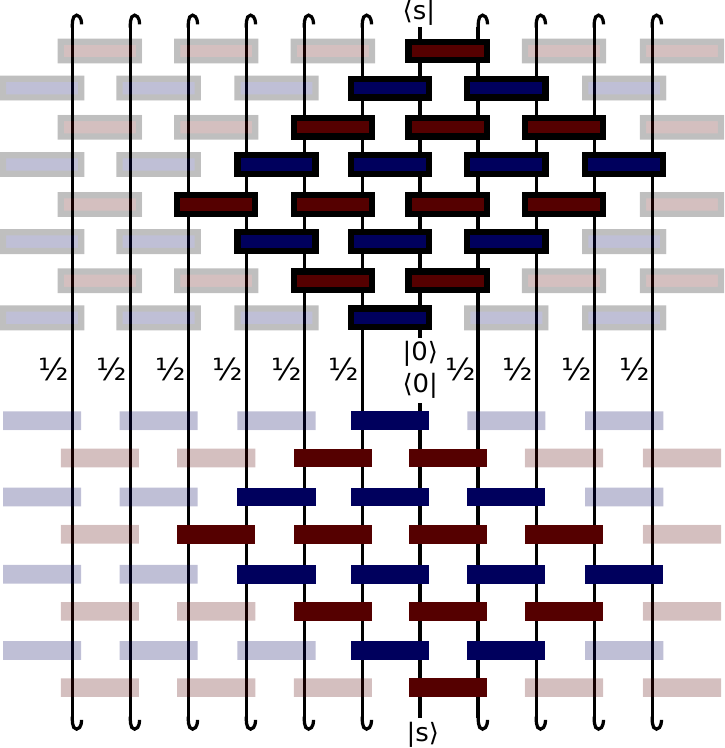}
 \caption{Diagrammatic representation of (\ref{eq:unitary rhomboid value}) for $t=4$. Unitaries outside of lightcones cancel and two rhomboids remain. A larger system results in more empty traces that do not contribute as they each have a factor $1/2$ attached. Sites are shown in the horizontal direction, unitaries $U_i$ ($V_i$) are shown as blue (red) boxes. Their daggered counterparts lack a thick border.}
 \label{fig:lightcones}
\end{figure}

The evaluation of $R(U,s)$ can be sketched diagrammatically as in Fig.~\ref{fig:lightcones}. Unitaries outside the lightcones cancel in pairs with their daggered counterparts and two rhomboids of width $2t+1$ sites remain. Considering only this part of the chain, and evaluating the diagram timestep by timestep, starting from $\rho_0$ in the middle, we encounter objects of dimension $4^{2t+1}$.

\begin{figure}
 \centering
 \includegraphics[width=6cm]{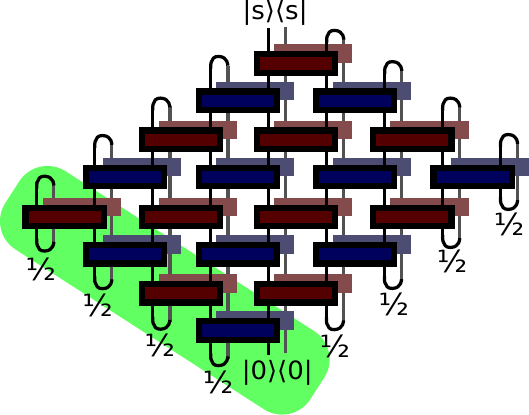}
 \caption{The rhomboids from Fig.~\ref{fig:lightcones} can be folded above each other. The diagram is contracted diagonally, beginning with the shaded green part.}
 \label{fig:folded rhomboids}
\end{figure}

After folding the daggered rhomboid upwards (Fig.~\ref{fig:folded rhomboids}), we can evaluate the folded rhomboids diagonally. This leads to a square root improvement, we encounter objects of dimension $2^{2t}$.
Note that this idea may be more generally applicable in tensor network contractions.

\begin{figure}
 \centering
 \includegraphics[width=7cm]{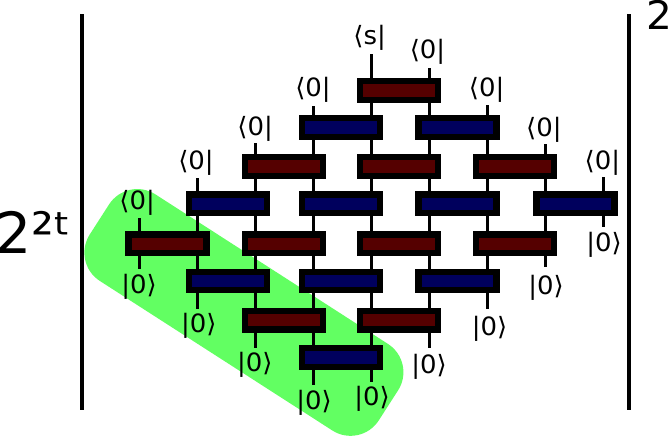}
 \caption{The absolute value squared of a single rhomboid. The diagram is contracted diagonally starting with the shaded green part. After averaging the unitaries, this diagram has the same value as Fig.~\ref{fig:folded rhomboids}. See section~\ref{sec:unitary details numeric} and appendix~\ref{sec:unitary appendix u-turns} for details.}
 \label{fig:single rhomboid}
\end{figure}

Owing to the single-site Haar invariance, the average of (\ref{eq:unitary rhomboid value}) remains the same when replacing all of the identities (``U-turns'') in the folded rhomboid diagram (Fig.~\ref{fig:folded rhomboids}) by $\ket{0}\!\bra{0}$. How this can be achieved is explained in detail in appendix~\ref{sec:unitary appendix u-turns}. We obtain two (disconnected) rhomboids that correspond to the absolute square of a single rhomboid as illustrated in Fig.~\ref{fig:single rhomboid}.  Again evaluating diagonally, we gain another square root as the objects only have dimension $2^t$.

When sampling $\avg{R(U,0)}_U$ according to the single-rhomboid method (Fig.~\ref{fig:single rhomboid})  we observe a higher variance than using the folded-rhomboids procedure (Fig.~\ref{fig:folded rhomboids}).
The data in Fig.~\ref{fig:unitary haar alpha} were compiled with the folded-rhomboids procedure only such that the strongly decreasing variance allows us to resolve the exponential decay of $\alpha(t)$. The exponentially decreasing variance is also indicative of self-averaging of (\ref{eq:unitary rhomboid value}).
To compile the data in Fig.~\ref{fig:unitary alphas}, we want to access longer times and therefore make use of the single-rhomboid optimisation. To counteract the increasing variance, we use an importance sampling technique.

To perform importance sampling, we extend the random variable set to include $s$ alongside $\{U_i\},\{V_i\}$. Then we generate samples according to the probability distribution $R(U,s)$ with a Metropolis algorithm. Now note that $\avg{R(U,0)}_U + \avg{R(U,1)}_U = 1$ follows immediately from (\ref{eq:unitary rhomboid value}). Thus the normalization of the probability distribution $R(U,s)$ is trivial. The average value of $\delta_{s0}$ with respect to this probability distribution therefore results in $\avg{R(U,0)}_U = (\alpha+1)/2$.

The method presented here allows us to reduce the complexity of calculating the time evolution of $t$ steps in a system of $2t+1$ sites (size of lightcone). Naively, time and space complexity both scale as $4^{2t+1}$. Our simplifications give two square roots improvement, yielding a scaling of $2^t$. Apart from the average over random unitaries, the numerical procedure is free of approximations.

The Monte Carlo aspect of the method can be generalized to improve the variance of expectation values $R_O = \avg{\Tr(O\psi_t)}$ of arbitrary observables $O$ over arbitrary ensembles of initial states or time evolutions determining $\psi_t$. Towards this end, extend $s$ to a POVM including $O$ instead of just $\dyad{0}{0}$ and $\dyad{1}{1}$ as for $R(U,s)$ above. Then perform Metropolis sampling of $\avg{\Tr(s\psi_t)}$ with respect to the random variables determining $\psi_t$ as well as $s$, which is taken as an additional random variable. Because of the normalization of the POVM, $\avg{\delta_{sO}}_{s,\psi_t} = R_O$.

\section{Conclusion and Outlook}

In this paper, we have studied one dimensional particle chains under a random unitary time evolution operator consisting of random nearest-neighbor gates. In spirit of Floquet evolution, the operator is repeated identically for subsequent timesteps.

We considered two cases, where the time evolution operator is a Gaussian circuit or consists of general unitaries. First, we were able to show strong results about the average evolution of chains of fermions under the Gaussian circuit time evolution. For Gaussian circuits inhomogeneous in space, we find that any initial state with vanishing two-point correlations at non-zero distances is simply scaled further towards the thermal mixture (\ref{eq:fermion inhomogeneous result}) and the initial two-point correlations can be recovered measuring expectation values; time evolution is localising. If the random time evolution operator is taken homogeneous in space, it delocalizes and leads to thermalization in the thermodynamic limit (\ref{eq:fermions homogeneous result}). We expect one can generalize our results to higher order correlation functions than the two-point functions studied in this work.

Next, we also considered spin chains under random unitary nearest-neighbor Floquet dynamics, inhomogeneous in space, with fixed finite local Hilbert space dimension.
Our main result is (\ref{eq:unitary depolarising channel}): On a single site, the average evolution acts as a depolarising channel, completely independent of any other initial sites.

We employ new numerical methods (section~\ref{sec:unitary details numeric}) to demonstrate that a time evolution composed of Haar distributed unitaries thermalizes. Under a different distribution with tunable random coupling strength, we find two regions of thermalization (strong coupling) and many-body localization (weak coupling), respectively.

As we have studied spins and fermions, it is natural to ask about a bosonic version  of the problem. Since, contrary to fermions, each bosonic mode defines an infinite-dimensional Hilbert space, the generalization of Haar unitaries may pose mathematical problems. Nevertheless, for future work it is conceivable to work directly in the symplectic space (that corresponds to the covariance matrices) which is finite.

Both our analytical results as well as the numerical method can readily be generalized to higher dimensions.
In the future, they may further also be applied to circuits with different topology and to Hamiltonian Floquet or stroboscopic dynamics with an ensemble of Hamiltonians having single-site Haar invariance.

\begin{acknowledgments}
During preparation of this manuscript, related work \cite{Amos_Andrea_2} appeared on arXiv that provides evidence for an MBL transition in a different unitary circuit with random coupling strength.

This project has received funding from the European Research Council (ERC) under
the European Union's Horizon~2020 research and innovation programme through the
ERC Starting Grant WASCOSYS (No.~636201), the ERC Consolidator Grant GAPS
(No.~648913), and the ERC Advanced Grant QENOCOBA (No.~742102). D.P.G.~acknowledges financial support of Severo Ochoa project SEV-2015-556 funded by MINECO.
\end{acknowledgments}

\appendix

\section{Gaussian circuits: Uncoupled case}
\label{sec:fermion uncoupled}

In this appendix, we calculate (\ref{eq:fermion homogeneous ctkN special values}), which we repeat for convenience:
\begin{equation}
 c(\tavg,k,N) = \frac{1}{4}\ \text{for}\ \frac{2\pi}{N/2} k = 0,\pi,2\pi.
 \nonumber
\end{equation}
For these values of $k$, $\hat{G}_k$ is real and may be absorbed by the Haar-invariant transformation $Q\to \hat{G}_k^\dagger Q\hat{G}_k$ in (\ref{eq:fermion homogeneous Ok}). In turn, $Q$ can be absorbed by the transformation $P\to Q^\dagger P$. Then we have simply $\hat{O}_k = P$.

This corresponds to two sites in the inhomogeneous uncoupled case where $Q_i = \Id_4$ in the time evolution operator (\ref{eq:fermion evolution operator}) and only $P_i$ are independently random. We find much stronger localization (intuitively, information cannot spread) where the constant $c(t\text{-avg})$ is one quarter:
\begin{equation}
 \avg{\Gamma_{t\text{-avg}}} = \frac{1}{4}\Gamma_0.
 \label{eq:fermion uncoupled}
\end{equation}

To show this it suffices to consider the first two sites $\Gamma_0^{1,2}$ and $P_1\in O(4)$. We introduce an arbitrary $A\in O(4)$ by $P_1 \to AP_1 A^\dagger$ using Haar invariance
\begin{equation}
 \avg{\Gamma^{1,2}_t} = \avg{A P_1^t A^\dagger \Gamma_0^{1,2} A P_1^{t\dagger} A^\dagger}
\end{equation}
and are free to integrate $A$ over the orthogonal group. The integral can be evaluated \cite{Weingarten_orthogonal} as
\begin{align}
 \avg{\Gamma^{1,2}_t} &= \frac{1}{12}\left\avg{(\Tr P_1^t)^2 - (\Tr P_1^{2t})\right}\Gamma_0^{1,2}
 \label{eq:fermion uncoupled integrated} \\
 &= \frac{1}{12}\left\avg{\left(\sum_{i=1}^4 e^{i\beta_i t}\right)^2 - \sum_{i=1}^4 e^{i\beta_i 2t} \right}\Gamma_0^{1,2}, \label{eq:fermion uncoupled evs}
\end{align}
which is determined by the spectrum $\{e^{i\beta_i},i=1,2,3,4\}$ of $P_1$.
We can evaluate this in the time average by observing that almost always 
\begin{equation}
\beta_1 = -\beta_2,\ \beta_3 = -\beta_4\ \text{for}\ \det P_1=+1
\end{equation}
and
\begin{equation}
\beta_1 = -\beta_2,\ \beta_3 = 0, \beta_4=\pi\  \text{for}\ \det P_1 = -1.
\end{equation}
One can then show that in the time average of (\ref{eq:fermion uncoupled evs}), $\avg{\cdot}_{\det P_1 =+1} = 4 - 0$ and $\avg{\cdot}_{\det P_1=-1} = 4 - 2$. Altogether the prefactor in (\ref{eq:fermion uncoupled integrated}) matches the $1/4$ announced in (\ref{eq:fermion uncoupled}).

\section{Spins: Uncoupled case}
\label{sec:unitary appendix uncoupled}
In this appendix, we find $\alpha(t)$ for the completely uncoupled probability distribution (\ref{eq:unitary tunable}), the limit $h=0$.
In that case, $U_i=u_{i,L}\otimes u_{i,R}$ and $V_i = v_{i,L}\otimes v_{i,R}$ are tensor products of single-site unitaries from the $U(2)$ Haar distribution.
It then suffices to consider only one site $\rho_0 = \Id_2/2 + \bar{\rho}_0$ as all sites are completely independent. Using the transformation $u\to v^\dagger u$, the $v$ can be Haar-absorbed into the $u$, and we have the time evolution $\avg{\rho_t} = \avg{u^t\rho_0u^{t\dagger}}$ which we evaluate for general dimension of $\rho_0$ and $u$ \cite{ZnidaricPRL2011,Gessner_Breuer}.

By Haar invariance the transformation $u\to w^\dagger u w$ shows
\begin{equation}
 \avg{\rho_t} = \avg{ w^\dagger u^t w \rho_0 w^\dagger u^{t\dagger} w }.
\end{equation}
We can integrate out $w$ with formula (\ref{eq:haar 2-design}) and get the result
\begin{equation}
 \avg{\rho_t} = \frac{\Id_d}{d} + \frac{\lambda(t)-1}{d^2-1}\bar{\rho}_0
\end{equation}
with the spectral form factor
\begin{equation}
 \lambda(t) = \left\avg{\Tr u^t \Tr u^{t\dagger}\right},
\end{equation}
which is just (\ref{eq:unitary details depolarising channel}) and (\ref{eq:unitary lambda}) for a one-site chain and empty set $I$.
For Haar-distributed $u\in U(d)$, the spectral form factor saturates at its maximal value $\lambda(t) = d$ for $t\geq d$ \cite{Mehta}.
In particular, for our $d=2$ chain and $t>1$, $\lambda(t) = 2$ and the final state (\ref{eq:unitary depolarising channel}) stays constant with $\alpha = 1/3$.

\section{Spins: Simplification for numerical calculations}
\label{sec:unitary appendix u-turns}

In this section, we show that the computationally more efficient single rhomboid contraction in Fig.~\ref{fig:single rhomboid} is equivalent to the function $R(U,s)$ from equation~(\ref{eq:unitary rhomboid value}) when taking the average in $U_i,V_i$. This is needed in section~\ref{sec:unitary details numeric}.

As argued in section~\ref{sec:unitary details numeric}, $R(U,s)$ is equal to the folded rhomboids in Fig.~\ref{fig:folded rhomboids}. Let us consider each site (i.e.~column) of that diagram in turn, apart from the central site containing $s$. The identities (``U-turns'') at the top and bottom of the column can be expanded. Linearity gives four new diagrams with all combinations of $\dyad{0}{0}$ and $\dyad{1}{1}$. Each column has either $U_i$ or $V_i$ both at the top and bottom. In the first case, single-site Haar invariance 
\begin{equation}
U_i\to (w\otimes\Id_2)U_i, V_i\to(\Id_2\otimes w^\dagger)V_i
\end{equation}
allows to insert $w$ and $w^\dagger$ that cancel everywhere except at the very top. By choosing $w=\sigma_x$ the Pauli matrix, $\dyad{1}{1}$ at the top can be transformed into $\dyad{0}{0}$. This process can be repeated similarly to transform a $\dyad{1}{1}$ at the bottom into $\dyad{0}{0}$.

In total, applying this procedure at all sites results in $4^{2t}$ identical diagrams where all top and bottom parts are $\dyad{0}{0}$ (except for $\dyad{s}{s}$). This only partially cancels with $2^{-2t}$ from the original bottom ``U-turns'', giving the prefactor. The two rhomboids of unitaries and daggered counterparts are then disconnected and can be written as the absolute square of a single rhomboid.

\nocite{Edelman_Rao_RMT,MPU}
\bibliography{randomevolutionBibliography}

\end{document}